\documentclass[12pt]{article}
\usepackage{epsfig}

\def\figsizeB{6.4}
\def\figsizeC{6.4}
\def\figure#1#2#3{\epsfxsize=#3truein
\centerline{\epsffile{fig_#1.eps}}
\centerline{\vbox{\bigskip \bigskip {\bf \noindent Figure #1.} #2}}
\bigskip }
\def\mres{m_{res}}

\def\mf{m_f}
\def\m0{m_0}
\def\Ls{L_s}
\def\ainv{a^{-1}}
\def\mev{\ \rm{MeV} \ }
\def\gev{\ \rm{GeV} \ }

\def\Psibar{\overline{\Psi}}

\def\Chi{X}
\def\Chibar{\overline{X}}

\def\slash{\!\!\!\!/}

\def\spose#1{\hbox to 0pt{#1\hss}}
\def\ltapprox{\mathrel{\spose{\lower 3pt\hbox{$\mathchar"218$}}
 \raise 2.0pt\hbox{$\mathchar"13C$}}}
\def\gtapprox{\mathrel{\spose{\lower 3pt\hbox{$\mathchar"218$}}
 \raise 2.0pt\hbox{$\mathchar"13E$}}}
\def\inapprox{\mathrel{\spose{\lower 3pt\hbox{$\mathchar"218$}}
 \raise 2.0pt\hbox{$\mathchar"232$}}}
\def\one{The quenched GDWF nucleon (top points) and the rho (bottom points)
mass in lattice units vs. $\mf$ at $\Ls = 16$ for three values of
$\beta$. The extrapolated $\mf=0$ rho mass is used to set the scale.
}
\def\two{Matching the scale of quenched GDWF with the scale of
quenched DWF. The nucleon (top points) and the rho (bottom points)
mass in lattice units for three values of $\beta$ is shown.  The
diamonds are for DWF from reference \cite{columbia_quenched_dwf} and
the crosses are for GDWF. Also the $\beta$ values for DWF are the
bottom numbers of the x-axis while the GDWF $\beta$ values are the top
numbers of the x-axis.
}
\def\three{The inverse lattice spacing in $\gev$ vs. the quenched GDWF
$\beta$ values used n this work.
}
\def\four{The ten smallest magnitude eigenvalues of $H_4(m_0)$
vs. $\m0$. An aggregate of the results from 20 independent
configurations is plotted in each plot. The left column is from
0-flavor Wilson (quenched) simulations while the right column is from
2-flavor Wilson with mass $-\m0$. Horizontally the 0-flavor and
2-flavor $\beta$ values correspond to the same lattice spacing (Figure
2). From top to bottom $\ainv \approx 1.0 \gev$, $1.4 \gev$ and $2.0
\gev$. Notice the difference in the y-axis scale for the different 
lattice spacings. 
}
\def\five{The 100 smallest magnitude eigenvalues of $H_4(m_0)$ at
$\m0=1.9$ were measured for 110 independent configurations for
$\beta=5.85$ and 0-Wilson flavors and for $\beta=4.6$ and 2-Wilson
flavors. Both have $\ainv \approx 1.4 \gev$.  The fraction of
eigenvalues with values between $\lambda$ and $\lambda + d\lambda$ is
plotted vs. $\lambda$. Here $d\lambda \approx 0.007$.
}
\def\six{The pion (diamonds), rho (squares) and nucleon (stars)
masses vs. $\ainv$ in $\gev$ measured using 2-flavor dynamical Wilson
fermions (both for the propagator and the sea quarks) with mass
$\mu=-1.9$. The scale is set using the GDWF rho mass. The straight line
marks the cutoff. Masses above that line are above the lattice cutoff.
}
\def\seven{The expectation value of the $\Psibar \gamma_5 \Psi$
condensate in lattice units vs. $\ainv$ in $\gev$ using 2-flavor
dynamical Wilson fermions (both for the propagator and the sea quarks)
with mass $\mu=-1.9$. It is consistent with zero well within the
error.  The lattice spacing is set using the GDWF $\rho$ mass. The
straight line marks the y-axis zero line.
}
\def\eight{The quenched DWF and GDWF residual masses $\mres$ vs. $L_s$
at $\ainv \approx 1.0 \gev$. The top points (squares) are from
quenched DWF simulations at $\beta=5.7$ while the bottom points
(diamonds) are from quenched GDWF simulations at $\beta=4.4$. In both
cases $\m0=1.9$ and $\mres$ was measured at $\mf=0.02$.
}
\def\nine{The quenched DWF and GDWF residual masses $\mres$ vs. $L_s$
at $\ainv \approx 1.4 \gev$. The top points (squares) are from
quenched DWF simulations at $\beta=5.85$ while the bottom points
(diamonds) are from quenched GDWF simulations at $\beta=4.6$. In both
cases $\m0=1.9$ and $\mres$ was measured at $\mf=0.02$.
}
\def\ten{The quenched DWF and GDWF residual masses $\mres$ vs. $L_s$
at $\ainv \approx 2.0 \gev$. The top points (squares) are from
quenched DWF simulations at $\beta=6.0$ while the bottom points
(diamonds) are from quenched GDWF simulations at $\beta=4.8$. In both
cases $\m0=1.9$ and $\mres$ was measured at $\mf=0.02$.
}
\def\eleven{The quenched GDWF residual mass $\mres$ vs. $m_f$ at
$L_s=16$ for three values of $\beta$.
}
\def\twoelve{The pion mass squared vs. $\mf$ from quenched GDWF
simulations. The squares are the measured data points, the straight
line is a least $\chi^2$ fit and the star is the $\mf=0$ extrapolated
point. Here $\beta=4.4$ corresponding to $\ainv \approx 1.0 \gev$,
$\m0=1.9$ and $L_s = 16$.  
}
\def\thirteen{The pion mass squared vs. $\mf$ from quenched GDWF
simulations. The squares are the measured data points, the straight
line is a least $\chi^2$ fit and the star is the $\mf=0$ extrapolated
point. Here $\beta=4.6$ corresponding to $\ainv \approx 1.4 \gev$,
$\m0=1.9$ and $L_s = 16$.
}
\def\fourteen{The pion mass squared vs. $\mf$ from quenched GDWF
simulations. The squares are the measured data points, the straight
line is a least $\chi^2$ fit and the star is the $\mf=0$ extrapolated
point. Here $\beta=4.8$ corresponding to $\ainv \approx 2.0 \gev$,
$\m0=1.9$ and $L_s = 16$.
}

\begin{document}

\title{\bf Gap Domain Wall Fermions}
\vskip 1. truein

\author{Pavlos M.\ Vranas \\
IBM T.\ J.\ Watson Research Center \\
Yorktown Heights, NY 10598, USA \\
vranasp@us.ibm.com}

\maketitle

\begin{abstract}
I demonstrate that the chiral properties of Domain Wall Fermions (DWF)
in the large to intermediate lattice spacing regime of QCD, 1 to 2
GeV, are significantly improved by adding to the action two standard
Wilson fermions with supercritical mass equal to the negative DWF five
dimensional mass. Using quenched DWF simulations I show that the
eigenvalue spectrum of the transfer matrix Hamiltonian develops a
substantial gap and that the residual mass decreases
appreciatively. Furthermore, I confirm that topology changing remains
active and that the hadron spectrum of the added Wilson fermions is
above the lattice cutoff and therefore is irrelevant. I argue that
this result should also hold for dynamical DWF and furthermore that it
should improve the chiral properties of related fermion methods.
\end{abstract}

\newpage

%%%%%%%%%%%%%%%%%%%%%%%%%%%%%%%%%%%%%%%%%%%%%%%%%%%%%%%%%%%%%%%%%%%%%%%%%%%%%%%
\section{\label{sec:intro} Introduction}
%%%%%%%%%%%%%%%%%%%%%%%%%%%%%%%%%%%%%%%%%%%%%%%%%%%%%%%%%%%%%%%%%%%%%%%%%%%%%%%

Domain Wall Fermions (DWF) and other closely related methods provide
the most faithful lattice regularization of QCD with unprecedented
chiral symmetry and topological properties. These methods have
produced an impressive wealth of results and have brought about a new
and brave era for lattice gauge theory. Several proposals for
improvement have also been demonstrated.  The reader is referred to
the reviews 
\cite{vranas_dubna_rev, lat00_rev, fleming_phd_thesis, lat01_rev, lat02_rev, lat03_rev, lat04_rev, lat05_rev, general_rev},
related works \cite{related_works}, and references therein for more details. 
Here I will focus on DWF and QCD. The method I will described 
should apply to all closely related lattice fermions as well as to 
other gauge theories similar to QCD (such as ${\cal N}=1$ Super Yang-Mills 
\cite{fleming_kogut_vranas_sym}). The method was first proposed in 
\cite{vranas_dubna_rev}.

Using the new generation supercomputers and numerical simulation
algorithms one can now simulate dynamical QCD at zero and finite
temperature in the strong to intermediate coupling regime with lattice
spacing $a$ in the region $1 \gev \ltapprox \ainv \ltapprox 2 \gev$ and for
number of lattice points $L_s$ along the fifth dimension in the
neighborhood of $L_s \approx 20 \ \rm{to} \ 100$ depending on the
lattice spacing $a$. These $L_s$ values are now certainly within the
reach of the latest supercomputers. However, one obviously would like
to do better.  A method that would improve DWF so that smaller values
of $L_s$ are needed would be very welcomed.  And not only for the
obvious benefits in computing time but also as a matter of theoretical
principle. The reasons behind the need for rather large values of
$L_s$ are as involved as they are interesting. But they are also a
lattice artifact.  It should be possible to remove it without spoiling
the important features of DWF.

In this work I demonstrate that the chiral properties of DWF in the large
to intermediate lattice spacing regime $1 \gev \ltapprox \ainv
\ltapprox 2 \gev$ are significantly improved by adding to the action
two standard Wilson fermions with mass equal to the negative DWF five
dimensional mass $m_0$ (Wilson supercritical region). These fermions
produce a fermion determinant that is identically zero when the fifth
dimension transfer matrix Hamiltonian $H_4(m_0)$ has a
zero. Therefore, small eigenvalues of $H_4(m_0)$ ought to be
suppressed in the simulation. And it is precisely these eigenvalues
that are a lattice artifact and are directly responsible for the large
$L_s$ requirement. Suppressing them allows one to achieve same chiral
symmetry restoration for smaller values of $L_s$. Indeed, I show that
the eigenvalue spectrum of $H_4(m_0)$ develops a substantial {\bf gap}
and that the residual mass $\mres$, quantifying the remnants of chiral
symmetry breaking, decreases appreciatively. I use this key property
to name this fermion regulator as Gap Domain Wall Fermions (GDWF). For
other DWF related fermions (for example overlap-Neuberger fermions)
I argue that this method of adding gap fermions (GF) should also 
improve their chiral properties.

In order to make sure there are no adverse ``side-effects'', I confirm
that local topology changing (measured using the overlap index method)
remains active; instantons and anti-instantons with size larger than a 
lattice spacing are active in the gauge field configurations. 
Also, I confirm that the hadron spectrum of the added
Wilson fermions is above the lattice cutoff and therefore is
irrelevant. Furthermore, I check that parity is not broken and
therefore the added Wilson fermions are well outside the Aoki
phase. All numerical simulations in this work were performed on
512-node IBM BlueGene/L supercomputers.

This paper is organized as follows: In section \ref{sec:dwf}, DWF and
the difficulties with chiral restoration are briefly reviewed. In
section \ref{sec:gdwf}, the Gap Domain Wall Fermion (GDWF) method is
presented. In section \ref{sec:results}, numerical results using GDWF
in the quenched approximation on $16^3 \times 32$ lattices are given
for three lattice spacings $\ainv \approx 1.0$, $1.4$ and $2.0 \gev$, and are
compared with the quenched approximation of standard DWF at the same
lattice spacings. Also, in this section I argue that these results
will hold for dynamical GDWF as well. In section \ref{sec:pions}, an
example of a quenched two-flavor QCD simulation with GDWF at the
rather large lattice spacing $\ainv \approx 1.4\gev$ is presented. It is shown
that the pion mass acquires physical values already for an $L_s$ as
small as $24$ with $\mres$ contributing about $10 \%$ to the bare
quark mass.  In section \ref{sec:topology} I discuss issues that relate 
to topology. In section \ref{sec:costs}, the ease of implementing GDWF,
as well as the small GDWF computational cost added to the cost of a
DWF simulation is discussed.  In section \ref{sec:thoughts}, some
curious theoretical thoughts brought up by GDWF are pondered upon. In
section \ref{sec:explore}, I discuss the applicability of the method to
overlap and Neuberger fermions and furthermore to the ${\cal N}=1$ 
Super Yang-Mills theory. Also, I present open issues for
further exploration.  The conclusions are given in section
\ref{sec:conclusions}.

%%%%%%%%%%%%%%%%%%%%%%%%%%%%%%%%%%%%%%%%%%%%%%%%%%%%%%%%%%%%%%%%%%%%%%%%%%%%%%%
\section{\label{sec:dwf} DWF and the chiral restoration problem}
%%%%%%%%%%%%%%%%%%%%%%%%%%%%%%%%%%%%%%%%%%%%%%%%%%%%%%%%%%%%%%%%%%%%%%%%%%%%%%%

Lattice DWF \cite{kaplan_dwf, overlap, furman_shamir_dwf, vranas_schwinger_dwf, neuberger_fermions}
are defined in five dimensions. The fifth dimension has
$L_s$ lattice sites and the five-dimensional fermion has positive bare
mass $m_0$ (domain wall height).  The five dimensional Dirac operator
$D_{F}$ employs free boundary conditions at the edges of the fifth
dimension (walls). As a result the plus chirality fermionic components
are localized on one wall while the minus chirality components are
localized on the other.  The two chiralities are explicitly mixed with
a mass parameter $m_f$. The gauge fields are defined in four
dimensions only. They are the same along the fifth dimension and have
no fifth component. This allows for a definition of a transfer matrix
$T$ along the fifth direction that is the same in all ``slices'' along
that direction. The product of the transfer matrices along the fifth
direction is therefore $T^{L_s}$. The single particle Hamiltonian
$H_4(m_0)$ associated with this transfer matrix is then also
independent of the fifth dimension. It is defined in four dimensions
and, for the case where the fifth dimension is continuous, one can show
that $H_4(\m0) = \gamma_5 D\slash_w(-\m0)$ where $D\slash_w(-\m0)$ is
the standard Wilson fermion Dirac operator with mass $-\m0$. When the
fifth dimension is not continuous the Hamiltonian has a more
complicated form, but one can show that it has the same zero
eigenvalues as $H_4(\m0)$.

The five dimensional Dirac operator is given by the equation:
\begin{equation}
D_F(x,s; x^\prime, s^\prime; \m0, \mf) = \delta(s-s^\prime) D\slash(x,x^\prime; \m0)
+ D\slash^\bot(s,s^\prime; \mf) \delta(x-x^\prime; \mf)
\label{D_F}
\end{equation}
\begin{eqnarray}
D\slash(x,x^\prime; \m0) &=& {1\over 2} \sum_\mu\left[ (1+\gamma_\mu)
U_\mu(x) \delta(x+\hat\mu - x^\prime) + (1-\gamma_\mu)
U^\dagger_\mu(x^\prime) \delta(x^\prime+\hat\mu - x) \right] \nonumber \\
&-& (4 - m_0)\delta(x-x^\prime)
\label{Dslash_F}
\end{eqnarray}
\begin{equation}
D\slash^\bot(s,s^\prime; \mf) = \left\{ \begin{array}{ll} P_R
\delta(1-s^\prime) - m_f P_L \delta(L_s-1 - s^\prime) - \delta(0-
s^\prime) & s=0 \\ P_R \delta(s+1 - s^\prime) + P_L \delta(s-1 -
s^\prime) - \delta(s-s^\prime) & 0 < s < L_s-1 \\ -m_f P_R
\delta(0-s^\prime) + P_L \delta(L_s-2 - s^\prime) - \delta(L_s-1 -
s^\prime) & s = L_s -1
\end{array}
\right. 
\label{Dslash_perp_f}
\end{equation}
\begin{equation}
P_{R,L} = { 1 \pm \gamma_5 \over 2}
\label{g5_proj}
\end{equation}
where $\m0$ is the 5-dimensional mass representing the ``height'' of
the Domain Wall and $\mf$ is the explicit bare quark mass. The gamma 
matrices $\gamma$ are taken in the chiral basis.

The standard Wilson fermion Dirac operator is:
\begin{equation}
D\slash_w(x,x^\prime; -\mu) = D\slash(x,x^\prime; \mu)
\label{Dslash_wilson}
\end{equation}

The localization of the two chiral components on the opposite walls is
exponentially good.  The slowest decay coefficient is
proportional to the smallest, in absolute value, negative eigenvalue of
$H_4(m_0)$.  For infinite $L_s$ (overlap fermions) the two chiralities
completely decouple provided that $H_4(m_0)$ does not have eigenvalues
that are exactly zero. That possibility is of measure zero and is
therefore of no concern.  Nevertheless, at finite $L_s$, where
simulations are performed, the two chiralities will mix and break
chiral symmetry. This mixing is of a similar nature as the one produced
by a mass term. It is possible to calculate this ``effective'' mass
(usually called residual mass, $\mres$) and use it to quantify the
quality of the DWF regulator. Clearly at finite $L_s$ one would like
$H_4(m_0)$ to have a substantial gap which in turn would result to a
rapidly decreasing $\mres$ as $L_s$ is increased.

Obviously the spectrum of $H_4(\mu)$ at various values $\mu$ is of
particular importance. A small review is given below. It will serve to
explain GDWF.  The reader is referred to the original work \cite{overlap} 
for more details.  For any gauge field configuration $H_4(\mu)$ has the
same number of positive ($n_+$) and negative ($n_-$) eigenvalues for
$\mu < 0$.  However, as $\mu$ is increased above zero some eigenvalue
of $H_4(\mu)$ may cross zero and change sign.  Then $n_+ - n_-$ would
not be zero just after the crossing occurs. It has been shown that the
number and direction of crossings is directly related to the number of
instantons and anti-instantons present in the gauge configuration 
\cite{overlap, edwards_heller_narayanan_flow} and
that $n_+ - n_-$ is in fact equal, in a statistical sense, to the net (global)
topological charge of the gauge field configuration \cite{suri_pavlos_index}. 
The Atiyia-Singer index theorem is realized on the lattice in a statistical 
sense. These are rather remarkable properties.

A very nice way to look at the spectrum of $H_4(\mu)$ is to plot the
eigenvalues of $H_4(\mu)$ as a function of $\mu$.  This is an
eigenvalue flow diagram 
\cite{overlap, suri_pavlos_index, edwards_heller_narayanan_flow}
(see Figure 4). Instantons that are
larger than the lattice spacing are of course of physical interest and
it has been shown \cite{edwards_heller_narayanan_flow} that they 
produce crossings in rather
localized neighborhoods of $\mu$ that correspond to the edges of the
standard Brillouin zones. For a single flavor DWF one picks $\mu =
m_0$ in-between the first and second set of crossings. Since this is a
finite range no fine-tuning is required. In the continuum limit the
range for one flavor extends from $0$ to $2$ and the width of the
neighborhoods where crossings occur tends to zero.  The location and
width of the crossing neighborhoods is renormalized from their
continuum values as the coupling is made stronger. For example, for
lattice spacing $\ainv \approx 1.4 \gev$, the first set of crossings occur in
the neighborhood of $\mu_{\rm min} \approx 0.9$ and the second in the
neighborhood of $\mu_{\rm max} \approx 2.2$. Their width is 
approximately 0.2.  However, small instantons
of the size of the lattice spacing are generated/destroyed because
they can ``come up/fall in'' through the discrete lattice.  This
generates additional crossings throughout the 
$\mu_{\rm min}$, $\mu_{\rm max}$ region.

In a numerical simulation at small lattice spacing (weak coupling)
there are few to no such small instanton crossings.  The simulation is
performed at a $\mu=\m0$ in the middle of the relevant range. At that
value, since there are no crossings, the eigenvalue gap is large and
therefore the localization on the walls is good. As a result, the two
chiralities mix very weakly and break chiral symmetry minimally (as a
result $\mres$ decreases rapidly with increasing $L_s$).  However, at
large lattice spacings (strong coupling) the small instantons generate
crossings across the whole range and therefore also close to $\m0$. As
a result the eigenvalue gap becomes very small. To be sure the
probability of an eigenvalue crossing {\it exactly} at $\m0$ is
zero. There is always a gap but it may be very small at large lattice
spacings. The challenge is to suppress the crossings due to the
lattice spacing size instantons, which are obviously an artifact of
the lattice ``discreteness'', without destroying the crossings
due to the all-important physical instantons with size of many lattice
spacings.

%%%%%%%%%%%%%%%%%%%%%%%%%%%%%%%%%%%%%%%%%%%%%%%%%%%%%%%%%%%%%%%%%%%%%%%%%%%%%%%
\section{\label{sec:gdwf} Gap Domain Wall Fermions}
%%%%%%%%%%%%%%%%%%%%%%%%%%%%%%%%%%%%%%%%%%%%%%%%%%%%%%%%%%%%%%%%%%%%%%%%%%%%%%%

A method that would produce a sizeable gap in the spectrum of
$H_4(m_0)$ for lattice spacings in the $1 \gev \ltapprox \ainv \ltapprox 2 \gev$
region would be of great use. Here I describe such a method. Gap
Domain Wall Fermions are similar to standard Domain Wall
Fermions but they have the desired {\it gap}.

Since $H_4(\m0) = \gamma_5 D_w(-\m0)$ where $D_w(-\m0)$ is the
standard Wilson fermion Dirac matrix one can induce a larger gap by
adding to the theory standard dynamical Wilson fermions with mass
$-\m0$. Here I add two flavors. When integrated out these fermions 
contribute a factor of $det^2[D_w(-m_0)] = det^2[H_4(m_0)]$ to
the Boltzman weight.  Gauge field configurations for which $H_4(m_0)$
has small eigenvalues will be suppressed by this Boltzman weight and
therefore they will be sampled very infrequently. In particular, any
gauge field configuration for which $H_4(m_0)$ has a zero eigenvalue
is explicitly excluded (not to mention that the set of such
configurations is of measure zero).  A gap at $\m0$ is ensured. But
more to the point is the fact that this Boltzman weight
``repels'' gauge configurations for which the gap at $\m0$ is
small. One can expect a substantially larger gap even for strong
couplings. In section \ref{sec:results}, I present numerical results
that demonstrate that indeed this is the case. Notice that I have not
added any extra parameters since the Wilson fermions have mass equal
to the negative five dimensional mass which is already a parameter of
the theory.

The Wilson fermions that I added to the theory have mass $-\m0$ with
$m_0$ somewhere in the middle of the crossings region $[\mu_{\rm min},
\mu_{\rm max}]$. I have chosen $m_0 = 1.9$ which is a good choice for
the whole range of lattice spacings of interest. Such a mass is in the
supercritical region of Wilson fermion masses and is very heavy.  The
hadron spectrum, including the pions, of these two flavors of Wilson
fermions, should be above the cutoff. In that case their contribution to
the low energy physics of the theory is irrelevant. Indeed, in section
\ref{sec:results}, I present numerical results that demonstrate that
their hadron spectrum is above the cutoff. Furthermore, I show that 
parity is not broken and therefore the added Wilson fermions are 
well outside the Aoki phase.

As mentioned earlier, it is important that crossings due to 
the all-important physical instantons with size of many lattice
spacings are present. The added Wilson fermions have mass $-m_0$ and they
suppress the crossings around $m_0$ but have little effect further
away. Because $m_0$ is chosen somewhere in the middle of the allowed
range, the larger instanton crossings are not affected since they occur
at the edges of the allowed range. Again, in section \ref{sec:results},
I present numerical results that demonstrate that this is the case.

The GDWF full partition function for QCD is given by:
\begin{equation}
Z = \int [dU] \int [d\Chibar d\Chi] \int [d\Psibar d\Psi] \int [d\Phi^\dagger d\Phi] e^{-S}
\label{Z}
\end{equation}
$U_\mu(x)$ is the gauge field, 
$\Chi(x)$ is the 4-dimensional Wilson fermion field $\Psi(x,s)$ is the 5-dimensional fermion field, and
$\Phi(x,s)$ is the 5-dimensional bosonic Pauli Villars (PV) type field. 
$x$ is a coordinate in the 4-dimensional space-time box with extent $L$
along each of the directions, $\mu = 1,2,3,4$, and $s=0,1, \dots,L_s-1$, 
where $L_s$ is the size of the fifth direction and is taken to be an even 
number.  The action $S$ is given by:
\begin{eqnarray}
S =  S(\beta, L, L_s, m_0, m_f) &=&               \nonumber \\ 
& &  S_G(U; \beta, L)                             \nonumber \\ 
&+&  S_W(\Chibar, \Chi, U; -\m0, L)               \nonumber \\ 
&+&  S_F(\Psibar, \Psi, U, \m0, \mf, L, L_s)      \nonumber \\ 
&+&  S_{PV}(\Phi^\dagger, \Phi, U; \m0, L, L_s)
\label{action}
\end{eqnarray}
where: $S_G$, is the standard plaquette Wilson gauge action with
coupling $g$ ($\beta = 6 / g^2 $) \cite{wilson_original}. Of course,
any pure gauge action can be chosen instead.  In particular, one can
choose a lattice gauge action that most closely resembles the
continuum action.  For example, a popular choice is the Iwasaki action
\cite{iwasaki_action}. Because, as mentioned above, DWF maintain a
connection between their index and the topological charge, it is
possible to improve them by suppressing small instantons by an
appropriate choice of the lattice pure gauge action
\cite{vranas_dubna_rev, fleming_phd_thesis, columbia_iwasaki_imprv, other_gauge_imprv}.
The Iwasaki type actions have been shown to have this property but
mostly for lattice spacings in the lower end of the interval
considered here and below. In any case, in order to be able to
evaluate GDWF alone I did not want the improvements due to these
actions to obscure the results and that is why I use the standard
Wilson gauge action. However, it should be obvious that for production
simulations one should use an improved gauge action not only because
of the possible additional improvement on the chiral properties but
also because of the closer resemblance to the continuum gauge
action. $S_W$ is the standard Wilson fermion action
\cite{wilson_original} with mass $-\m0$. $S_F + S_{PV}$ is the
standard Domain Wall Fermion action with the Pauli Villars regulator,
five dimensional mass $\m0$ and explicit bare quark mass $\mf$
\cite{kaplan_dwf, overlap, furman_shamir_dwf, vranas_schwinger_dwf}.

%%%%%%%%%%%%%%%%%%%%%%%%%%%%%%%%%%%%%%%%%%%%%%%%%%%%%%%%%%%%%%%%%%%%%%%%%%%%%%%
\section{\label{sec:results} Numerical results}
%%%%%%%%%%%%%%%%%%%%%%%%%%%%%%%%%%%%%%%%%%%%%%%%%%%%%%%%%%%%%%%%%%%%%%%%%%%%%%%

In this section I present numerical results that demonstrate the
properties of GDWF.

Because of limited computational resources I use the ``quenched''
approximation for the DWF sea fermions. However, unlike standard
quenched simulations I consider large lattice spacings that are of
interest to dynamical DWF simulations.  I compare results obtained
from simulations with no dynamical Wilson flavors with results
obtained at the same lattice spacing with two dynamical Wilson flavors
with mass $-\m0$. I only use one value of $\m0=1.9$ throughout this
work. In order to compare results I match the lattice spacing between
the two cases by adjusting $\beta$. Here I achieve a $5\%$ or better
matching level at three values of the lattice spacing $\ainv \approx 1.0
\gev$, $1.4 \gev$ and $2.0 \gev$. Measurements are done using the DWF
operator at $\m0=1.9$. The space-time volume of all simulations is
$16^3 \times 32$. More technical details about the numerical
simulations are given in the appendix.

Therefore, the two cases correspond to standard quenched DWF
simulations and to quenched GDWF simulations. Again, the lattice
spacings I use are large and of interest to dynamical
simulations. I expect that the effects on chiral symmetry due to the
DWF quenching are to a large extent taken into account by simulating
at these large lattice spacings. To be sure the quenched approximation
does not allow for the calculation of a systematic error. However, if
previous results are of any guide I expect that quenching may affect
mass measurement results at the $10\%$ level. Since the improvements
observed here are at the one order of magnitude level I do not expect
that quenching is affecting the results in any significant way.

The first order of business is to match the lattice spacings between
DWF and GDWF.  I measure the $\rho$ and nucleon masses at $L_s=16$ and
extrapolate to $\mf=0$.  The results for three values of $\beta$ are
shown in Figure 1. These values of $\beta$ were chosen so that the
$\mf=0$ extrapolated $\rho$ masses match the quenched DWF masses of
\cite{columbia_quenched_dwf}.

The matching for the $\mf=0$ extrapolated masses is shown in Figure
2. The DWF $\rho$ masses match well (better than $5 \%$).  I choose
the DWF $\mf=0$ extrapolated $\rho$ mass to set the scale. The values
of $\beta$ shown correspond to $\ainv \approx 1.0 \gev$, $1.4 \gev$ and $2.0
\gev$. The nucleon masses also match well (better than $5 \%$) which
is an indication that the method works properly.  From now on I will
use these values of beta to perform comparisons of the chiral
properties of DWF and GDWF. The inverse lattice spacing in $\gev$
corresponding to the quenched GDWF $\beta$ values is plotted in Figure 3.

In Figure 4 the ten smallest magnitude eigenvalues of $H_4(m_0)$ are
plotted vs. $\m0$. The eigenvalues are calculated with an accuracy
$10^{-6}$ and are measured in $\m0$ steps of $0.025$. An aggregate of
the results from 20 independent configurations (separated by 20
configurations) is plotted in each plot. The left column is from
0-flavor Wilson simulations while the right column is from 2-flavor
Wilson with mass $-\m0$. Horizontally, the 0-flavor and 2-flavor
$\beta$ values correspond to the same lattice spacing (Figure 2). From
top to bottom $\ainv \approx 1.0 \gev$, $1.4 \gev$ and $2.0
\gev$. Notice the difference in the y-axis scale for the different
lattice spacings. The ``crosshairs'' indicate the $\m0=1.9$ point. One
can clearly see that the 2-flavor Wilson fermions generate a
substantial gap around $\m0$ where none existed before even at the
large lattice spacing $\ainv \approx 1.0 \gev$.

Furthermore, it is very important to observe in Figure 4 that the
2-flavor Wilson fermions generate the gap at a neighborhood of
$\m0=1.9$, but allow for a copious amount of crossings at the edges of
the allowed $\m0$ range.  As mentioned in section \ref{sec:dwf}, these
crossings correspond to instantons with size larger than a lattice
spacing and are of physical interest. Although Figure 4 shows the cumulative 
results of 20 configurations, by close inspection I confirmed that the number 
of crossings changes from configuration to configuration. This indicates
instanton, anti-instanton activity. However, with the current resolution 
I am not able to measure the net topological index. This is beyond the 
scope of this work. For further discussion see section \ref{sec:topology}.

In order to get a better picture of the small eigenvalue distribution
I chose to look at the $\m0=1.9$ ``cross-section'' for the 0 and 2
flavor cases at $\ainv \approx 1.4 \gev$ ($\beta = 5.85$, $ 4.6$) using a
larger number of eigenvalues and configurations. I measured the 100
smallest magnitude eigenvalues of $H_4(\m0)$ at $\m0=1.9$ for 110
independent configurations (separated by 20 configurations, i.e. a
total of 2200 configurations were generated). The fraction of
eigenvalues with values between $\lambda$ and $\lambda +
d\lambda$ is plotted vs. $\lambda$ in Figure 5. Here $d\lambda \approx
0.007$.  The difference in the distributions is telling. The 0-flavor
Wilson distribution has non-zero support at $\lambda = 0$ and is
raising almost linearly with a small slope as $\lambda$ increases.  On
the other hand, the 2-flavor Wilson distribution has zero support
between $\lambda = 0$ and $\lambda= 0.015$. After that it is raising very
slowly until $\lambda=0.05$.  It does not exceed the 0-flavor
distribution until $\lambda$ gets larger than $\approx 0.12$.  Larger
statistics are needed to paint a clearer picture. Of course it may be
that the distribution has zero support only at $\lambda=0$ and then it
increases perhaps quadratically. Nevertheless the point is that for a
numerical simulation of QCD that typically generates a few thousand
configurations the method creates a substantial gap.

As discussed, it is expected that the added 2-flavors of Wilson flavors
with supercritical mass $\mu=-1.9$ should have a hadron spectrum above
the lattice cutoff. This is verified in Figure 6. The pion
(diamonds), rho (squares) and nucleon (stars) masses vs. $\ainv$ in
$\gev$ measured using 2-flavor dynamical Wilson fermions (both for the
propagator and the sea quarks) is shown. The scale is set using the
GDWF $\rho$ mass. The straight line marks the cutoff. Masses above
that line are above the lattice cutoff. Clearly the Wilson hadron
spectrum is above the lattice cutoff. Most importantly the pion mass
is very heavy and above the cutoff.  This ensures that the added
2-flavors of Wilson fermions do not affect the low energy physics.  On
the other hand, they clearly affect the physics near the cutoff. For
example, they renormalize the value of $\beta$. At $\ainv \approx 1.4 \gev$,
$\beta$ changes from $5.85$ to $4.6$. But, of course, this is expected
and is of no consequence. 

Also, one may worry that the Wilson fermions may break parity since
they have mass in the supercritical region.  This is obviously not the
case as can be seen from Figure 4 where the corresponding operator
$H_4(\m0) = \gamma_5 D\slash_w(-\m0)$ has no zero eigenvalues at
$\m0=1.9$. Nevertheless, I measure the expectation value of $\Psibar
\gamma_5 \Psi$ for the three lattice spacings used in this work.  The
results are shown in Figure 7. Its value is of the order $10^{-6}$
with an error that is an order of magnitude larger. It is consistent
with zero well within the error. Clearly parity is not broken and one
is away from the Aoki phase \cite{Aoki}.

The residual mass $\mres$ is measured using the ratio method (see for
example \cite{columbia_quenched_dwf} ). In Figure 8 the quenched DWF
and GDWF residual masses $\mres$ vs. $L_s$ at $\ainv \approx 1.0 \gev$
are shown. The top points (squares) are from the quenched DWF
simulations at $\beta=5.7$, while the bottom points (diamonds) are from
the quenched GDWF simulations at $\beta=4.4$.  In both cases $\m0=1.9$
and $\mres$ was measured at $\mf=0.02$ (the value of $\mres$ is fairly
insensitive to the value of $\mf$ as can clearly be seen in Figure
11). The expected faster exponential decay and much smaller values of
$\mres$ are evident. Already at $\Ls=24$, $\mres$ is smaller by almost
an order of magnitude.  Furthermore, its value $\mres \approx 0.002$ is
rather nice for such a large lattice spacing. At $\Ls=40$ $\mres
\approx 0.0005$. The difference becomes more dramatic as the lattice
spacing is decreased.  The results for $\ainv \approx 1.4 \gev$ are shown in
Figure 9 and for $\ainv \approx 2.0 \gev$ are shown in Figure 10.  Figure 9
is most interesting since it corresponds to the lattice spacing of
dynamical $N_t=8$ thermodynamics at the critical temperature. At
$\Ls=24$ $\mres \approx 0.0006$ and at $\Ls=32$ $\mres \approx
0.0002$.

As mentioned above $\mres$ is rather insensitive to $\mf$. This is shown 
in Figure 11 for the GDWF simulations at all three values 
of $\beta$.

Regarding chiral symmetry, one of the most telling observables is the
pion mass.  At the end of the day any improvement method should result
to small pion masses.  In Figure 12 the GDWF pion mass squared
vs. $\mf$ from quenched GDWF simulations is shown. The squares are the
measured data points, the straight line is a least $\chi^2$ fit and
the star is the $\mf=0$ extrapolated point. Here $\beta=4.4$
corresponding to $\ainv \approx 1.0 \gev$, $\m0=1.9$ and $L_s =
16$. The straight line fit intersects the x-axis at $\mf \approx
-0.004$ which is consistent with the value of $\mres \approx 0.006$
from Figure 8. The data for $\beta=4.6$ corresponding to $\ainv
\approx 1.4 \gev$, are shown in Figure 13 and for $\beta=4.8$
corresponding to $\ainv \approx 2.0 \gev$ are shown in Figure 14.  In
both cases the straight line fit gives $m_\pi^2 \approx 0$ at $\mf =
0$ within the error bar.

%%%%%%%%%%%%%%%%%%%%%%%%%%%%%%%%%%%%%%%%%%%%%%%%%%%%%%%%%%%%%%%%%%%%%%%%%%%%%%%
\section{\label{sec:pions} Light pions in rugged terrain}
%%%%%%%%%%%%%%%%%%%%%%%%%%%%%%%%%%%%%%%%%%%%%%le%%%%%%%%%%%%%%%%%%%%%%%%%%%%%%%

In this section I present a simulation of quenched GDWF at $\ainv=
1.356(75) \gev$, $\Ls=24$ and $\mf=0.005$. The space-time volume is
$16^3 \times 32$ and $m_0 = 1.9$. Think of this as the icing on the
cake or the cherry on top of the sundae. The point is that this
simulation should closely resemble a dynamical GDWF simulation near
the $N_t=8$ critical point of QCD thermodynamics. Also notice that in
order to mimic the way dynamical simulations are analyzed I calculated
the lattice spacing using the $\rho$ mass at the bare quark mass
$\mf=0.005$ and not at the $\mf=0$ extrapolated value.

I find that $\mres = 0.00064(4)$ which is about $10\%$ of the explicit
quark mass $\mf=0.005$.  And finally the pion mass is under control. I
find $m_\pi = 140(40) \mev$ at $\mf=0.005$.

%%%%%%%%%%%%%%%%%%%%%%%%%%%%%%%%%%%%%%%%%%%%%%%%%%%%%%%%%%%%%%%%%%%%%%%%%%%%%%%
\section{\label{sec:topology} Topological issues}
%%%%%%%%%%%%%%%%%%%%%%%%%%%%%%%%%%%%%%%%%%%%%%%%%%%%%%%%%%%%%%%%%%%%%%%%%%%%%%%
The resolution of the flow diagrams of Figure 4 does not allow me to
measure the net index but does allow me to observe that the number and
location of crossings change from configuration to configuration.
This indicates the desired instanton anti-instanton activity. Measuring
the net index requires more computational resources and is outside the
scope of this paper. However, a few remarks are in order.

If the update algorithm ``smoothly'' transforms the gauge field
configuration then the net index will have to change in a smooth way
too. In that case the heavy Wilson fermion determinant will prohibit
any flow line from crossing through $\m0$ and as a result the net
index will not be able to change. The simulation will generate
configurations with the same net index as the initial configuration
and will not be able to tunnel between sectors. For example, for an
ordered start the net topology will always remain zero. This does not
change the ability of the simulation to generate crossings (for
example as in Figure 4). It simply means that the appearance /
disappearance of an instanton will always be accompanied by that of an
anti-instanton of some size at some location.

Because I could not measure the net index it is not clear if the HMC
Phi algorithm that I used in this work is capable of topologically
non-smooth gauge field evolution that would generate tunneling between
sectors.  However, this is an algorithmic issue and is not particular
to GDWF. For example, net index change is suppressed as the lattice
spacing gets smaller irrespectively of using or not using GDWF. The
gauge action barriers between topological sectors are a property of
QCD. This has not been identified as a problem yet because the
couplings used in today's simulations are not weak enough. Even so,
algorithms that are able to tunnel between sectors have already been
proposed (for example see \cite{top_alg} and references therein).

Therefore if net index change is important to the physics at hand one
should check the net index properties of HMC Phi under GDWF. If they
are not satisfactory then one should use a properly augmented
evolution algorithm. In any event although the subject of net topology
is of interest in many cases one only needs to stay within sector zero
provided that the volume is large for the physics at hand. For a very
nice investigation on the subject see \cite{LS}.

%%%%%%%%%%%%%%%%%%%%%%%%%%%%%%%%%%%%%%%%%%%%%%%%%%%%%%%%%%%%%%%%%%%%%%%%%%%%%%%
\section{\label{sec:costs} Algorithmic and computational costs}
%%%%%%%%%%%%%%%%%%%%%%%%%%%%%%%%%%%%%%%%%%%%%%le%%%%%%%%%%%%%%%%%%%%%%%%%%%%%%%

The algorithmic implementation of a new method is of course a matter
of effort and not a matter of concept. However, this effort is not
always small. A method that is simple and is easily implemented as an
extension of existing methods is highly desirable. This is the case
for GDWF. Any QCD code that employs standard DWF or related fermions
already has an implementation of the Wilson $D\slash_w(x,x^\prime; \mu)$ 
operator and evolution force term. Typically one should simply
have to add to the DWF force term the Wilson force term and to the
corresponding energy function the extra Wilson fermion energy.

The additional computational cost of adding 2-flavors of heavy Wilson
fermions to a 2-flavor dynamical DWF simulation is obviously at most 
an $1 /L_s$ fraction. For example, at $L_s=24$ it should be less than a
$\approx 5\%$ addition to the computation cost.

%%%%%%%%%%%%%%%%%%%%%%%%%%%%%%%%%%%%%%%%%%%%%%%%%%%%%%%%%%%%%%%%%%%%%%%%%%%%%%%
\section{\label{sec:thoughts} Curious thoughts}
%%%%%%%%%%%%%%%%%%%%%%%%%%%%%%%%%%%%%%%%%%%%%%le%%%%%%%%%%%%%%%%%%%%%%%%%%%%%%%

GDWF bring up a few thoughts that are worthwhile pondering about.

Clearly the addition of heavy fermions to the theory is an alteration
in the ultraviolet regime. But in the GDWF case it dramatically 
affects the infrared properties of the theory.  After all this
was its purpose. This is an example of physics above the cutoff
affecting the physics far below the cutoff. These ``spectator'' fermions
may be pointing to some interesting theoretical directions.

GDWF were described as the addition of two extra Wilson fermions with
supercritical mass. This leads to a fermionic determinant
that directly suppresses the unwanted gauge configurations. In this
light GDWF are simply a device for better simulations of QCD on the
lattice.  Perhaps there is nothing more to it than that. But it is
interesting that GDWF can be interpreted in two additional and
different ways:

Another way to think about GDWF is to consider the log of the two flavor
Wilson determinant (with mass $-\m0$) as part of the pure gauge
action. As mentioned above it is irrelevant and therefore it is a
valid addition to the lattice gauge action. Thinking about it this way
opens the method to be applied to the existing wealth of improvement 
techniques for DWF and related fermions.

Yet another and perhaps most curious way to think of GDWF is to extend
the five dimensional DWF Dirac operator to include two additional
diagonal terms along the fifth direction. These terms can be inserted
``behind'' the walls. The determinant of this new Dirac operator is
the product of the original DWF determinant times the two flavor
Wilson determinant. More specifically, one will have $\Ls+2$ fermion
fields (but still $\Ls$ Paulli-Villars fields) and the DWF Dirac
operator of eq. \ref{D_F} will have to be augmented to the GDWF
operator by extending it along the fifth dimension to include two
$D\slash$ terms (see eq. \ref{Dslash_F}) along the diagonal. In
matrix form on the fifth dimension index the GDWF operator is (for
simplicity I have set $\mf=0$, all blank entries are zero and 
$P_{R, L}$ is given in eq. \ref{g5_proj} ):

\begin{equation} \left( \begin{array}{ccccccc}
  D\slash   &              &              &              &              &              &            \\
            & D\slash - 1  & P_R          &              &              &              &            \\
            & P_L          & D\slash - 1  &              &              &              &            \\
            &              &              & \ddots       &              &              &            \\
            &              &              &              & D\slash - 1  & P_R          &            \\
            &              &              &              & P_L          & D\slash - 1  &            \\
            &              &              &              &              &              & D\slash           
\end{array} \right) 
\label{eq:gdwf_mat}
\end{equation}

In some sense this is a natural extension of DWF. One can think of
this as having fermions beyond the walls that do not communicate
directly with the fermions inside the walls. Their presence is felt
only through their coupling to the gauge field in the bulk. Notice
that the fifth direction mid-point reflection property of DWF is
maintained this way.  However, also notice that this way one adds two
heavy Wilson fermions per DWF flavor. For example, 2-flavor QCD will
have an additional four flavors of heavy Wilson fermions. In the
numerical simulations of this work I only used 2-flavors of Wilson
fermions.  However, using more flavors should not present problems.

%%%%%%%%%%%%%%%%%%%%%%%%%%%%%%%%%%%%%%%%%%%%%%%%%%%%%%%%%%%%%%%%%%%%%%%%%%%%%%%
\section{\label{sec:explore} Open issues}
%%%%%%%%%%%%%%%%%%%%%%%%%%%%%%%%%%%%%%%%%%%%%%%%%%%%%%%%%%%%%%%%%%%%%%%%%%%%%%%

As mentioned earlier one does not expect dynamical simulations of GDWF
to change the quenched results by much, simply because the lattice
spacings considered here are large and are the ones of current
interest to dynamical simulations. Nevertheless, dynamical tests of
GDWF would be very welcomed.

The 2-flavors of Wilson fermions that were added in the action are
designed to produce a determinant $det[H_4]^2$. This suppresses the
configurations for which the transfer matrix of DWF has eigenvalue
$1$. However, $H_4$ is the transfer matrix Hamiltonian for DWF with a
continuous fifth direction. Here I used DWF with a discrete fifth
direction. To be sure, the continuous and discrete transfer matrix
Hamiltonians have the same zeroes. But the continuous transfer matrix
Hamiltonian used here may be more effective for the overlap-Neuberger 
fermions \cite{overlap, neuberger_fermions} as well as 
their variants and improvements
(see \cite{vranas_dubna_rev, lat00_rev, fleming_phd_thesis, lat01_rev, lat02_rev, lat03_rev, lat04_rev, lat05_rev, general_rev} 
and references therein) since it directly relates to them. On the
other hand, for DWF with discrete fifth dimension one may want to use
instead of a standard Wilson fermion an augmented Wilson fermion that
will perhaps further improve the chiral properties of DWF. The exact
form of $H_4$ for the discrete case is known exactly\cite{overlap,
borici}. Both of these observations deserve further
scrutiny. Furthermore, investigating the applicability of GF to other
related fermion regulators \cite{other_related_fermions} is of great
interest.

Throughout this work the reader may have been wondering why one
should only consider two flavors of Wilson fermions. Why not more?
This is indeed a very interesting question deserving further
exploration for the obvious extra benefits it may provide in the cost
of numerical simulations, but also for purely theoretical reasons. Is a
theory with many spectator fermions above the cutoff of any interest?

As mentioned earlier, in production simulations GDWF should
be used with improved gauge actions (for example Iwasaki). It is not
clear if this will further improve the chiral properties. This
certainly deserves more investigation. However, it should be used
anyway because it more closely resembles the continuum gauge action.

As discussed in section \ref{sec:topology} the subject of net topology
change is beyond the scope of this work. However it is an interesting
open issue that needs further investigation. Besides the algorithmic
approach discussed in section \ref{sec:topology} one may want to
consider alternative ideas. For example, it has been proposed that
using ${D\slash_w}^2 + h^2$, where $h$ is a small real number, instead
of ${D\slash_w}^2$ for the two heavy Wilson flavors will produce more
frequent net-topology changes as $h$ is increased from zero
\cite{christ_private, vranas_dubna_rev}. Clearly $h$ interpolates between GDWF ($h=0$)
and DWF ($h >> 1$).

Finally, simulations of the ${\cal N} = 1$ Super Yang-Mills theory are
possible \cite{neuberger_fermions, kaplan_schmaltz_sym} and have been
performed \cite{fleming_kogut_vranas_sym} using DWF. Issues similar to
the chral symmetry of QCD are present and therefore one would expect
that GDWF should work there as well.  Further exploration is needed.

%%%%%%%%%%%%%%%%%%%%%%%%%%%%%%%%%%%%%%%%%%%%%%%%%%%%%%%%%%%%%%%%%%%%%%%%%%%%%%%
\section{\label{sec:conclusions} Conclusions}
%%%%%%%%%%%%%%%%%%%%%%%%%%%%%%%%%%%%%%%%%%%%%%%%%%%%%%%%%%%%%%%%%%%%%%%%%%%%%%%

In this work I demonstrated that the chiral properties of Domain Wall
Fermions (DWF) in the large to intermediate lattice spacing $a$ regime
of QCD, $1 \gev \ltapprox \ainv \ltapprox 2 \gev$, are significantly
improved by adding to the action two standard Wilson fermions with
supercritical mass equal to the negative DWF five dimensional mass
$\m0$. Here I used $\m0 = 1.9$.  I performed quenched DWF simulations
and showed that the eigenvalue spectrum of the transfer matrix
Hamiltonian develops a substantial {\bf gap} and that the residual
mass decreases appreciatively (by about an order of magnitude). I used
this key property to name this fermion regulator as Gap Domain Wall
Fermions (GDWF). For other DWF related fermions (for example
overlap-Neuberger fermions) I argued that this method of adding ``gap
fermions'' (GF) should also improve their chiral properties.

In order to make sure there are no adverse ``side-effects'', I confirmed
that local topology changing (measured using the overlap index method)
remains active; instantons and anti-instantons with size larger than a 
lattice spacing are active in the gauge field configurations. 
Also, I confirmed that the hadron spectrum of the added
Wilson fermions is above the lattice cutoff and therefore is
irrelevant. Furthermore, I checked that parity is not broken and
therefore the added Wilson fermions are well outside the Aoki
phase. All numerical simulations in this work were performed on
512-node IBM BlueGene/L supercomputers.

Furthermore, I argued that the results of this work should also hold
for dynamical GDWF since I considered rather large lattice spacings
which are typical to currently possible dynamical DWF simulations. In
particular, the middle of the range of the lattice spacings $1 \gev
\ltapprox \ainv \ltapprox 2 \gev$ approximately corresponds to the
$N_t=8$ critical coupling of the QCD thermal phase transition. This
makes the method very appealing for dynamical QCD thermodynamics.

%%%%%%%%%%%%%%%%%%%%%%%%%%%%%%%%%%%%%%%%%%%%%%%%%%%%%%%%%%%%%%%%%%%%%%%%%%%%%%%
\section*{Acknowledgments}
%%%%%%%%%%%%%%%%%%%%%%%%%%%%%%%%%%%%%%%%%%%%%%%%%%%%%%%%%%%%%%%%%%%%%%%%%%%%%%%
I would like to thank my friend George Fleming for many discussions in
the physics of QCD and the inner-workings of DWF. I would like to thank 
Prof. N. Christ for valuable discussions regarding topological properties. 
I would like to thank the QCDOC collaboration for providing me with the 
Columbia Physics Code (CPS). And of course many thanks to the IBM Blue 
Gene team for allowing me access to the BlueGene/L supercomputer at the 
IBM T.J. Watson Research Center at Yorktown Heights, NY.

%%%%%%%%%%%%%%%%%%%%%%%%%%%%%%%%%%%%%%%%%%%%%%%%%%%%%%%%%%%%%%%%%%%%%%%%%%%

%%%%%%%%%%%%%%%%%%%%%%%%%%%%%%%%%%%%%%%%%%%%%%%%%%%%%%%%%%%%%%%%%%%%%%%%%%%%%%%
\section*{Appendix}
%%%%%%%%%%%%%%%%%%%%%%%%%%%%%%%%%%%%%%%%%%%%%%%%%%%%%%%%%%%%%%%%%%%%%%%%%%%%%%%
The evolution in all simulations was done using the standard HMC Phi
algorithm. In all cases the trajectory length was set to 0.5. The step
size was set to 0.005 or 0.01 depending on the acceptance rate. The
acceptance rate in all cases was above $85\%$. An initial 200
trajectories were used for thermalization.  A total of 50 to 100
measurements were done for all propagators. All measurements were
separated by 20 trajectories. The conjugate gradient residual was set
to $10^{-7}$.

The 2-flavor Wilson simulations are obviously full dynamical
simulations in the Wilson fermions (they are quenched only with
respect to the DWF fermions). The conjugate gradient residual was set
to $10^{-6}$. The conjugate gradient iterations for the evolution
varied between 100 and 200.

All fits in this work including the propagator fits to extract the
hadron masses (not shown) had a $\chi^2$ per degree of freedom less
than $1$.

\vfil

%%%%%%%%%%%%%%%%%%%%%%%%%%%%%%%%%%%%%%%%%%%%%%%%%%%%%%%%%%%%%%%%%%%%%%%%%%%%%%%
\section*{Tables and Figures}
%%%%%%%%%%%%%%%%%%%%%%%%%%%%%%%%%%%%%%%%%%%%%%%%%%%%%%%%%%%%%%%%%%%%%%%%%%%%%%%
In this section the numerical simulation results are presented in the 
following tables and corresponding figures.

\clearpage

\begin{table}
\centering
\begin{tabular}[h]{||c|c|c|c||} \hline \hline
$\beta$	&  $m_f$       &  $m_\rho$   &  $m_N$      \\ \hline \hline
4.4     &  0.0         &  0.720(12)  &  0.969(23)  \\ \hline
4.4     &  0.02        &  0.759(10)  &  1.058(26)  \\ \hline
4.4     &  0.04        &  0.789(10)  &  1.115(14)  \\ \hline
4.4     &  0.06        &  0.815(11)  &  1.203(14)  \\ \hline
4.4     &  0.08        &  0.868(10)  &  1.275(12)  \\ \hline \hline
4.6     &  0.0         &  0.541(9)   &  0.712(25)  \\ \hline
4.6     &  0.02        &  0.596(7)   &  0.823(22)  \\ \hline
4.6     &  0.04        &  0.636(19)  &  0.935(20)  \\ \hline
4.6     &  0.06        &  0.690(6)   &  1.036(14)  \\ \hline
4.6     &  0.08        &  0.753(7)   &  1.154(15)  \\ \hline \hline
4.8     &  0.0         &  0.399(16)  &  0.558(30)  \\ \hline
4.8     &  0.02        &  0.441(27)  &  0.640(47)  \\ \hline
4.8     &  0.04        &  0.522(9)   &  0.776(19)  \\ \hline
4.8     &  0.06        &  0.589(7)   &  0.890(9)   \\ \hline
4.8     &  0.08        &  0.644(6)   &  0.993(11)  \\ \hline \hline
\end{tabular}
\caption{
The $\rho$ ($m_\rho$) and nucleon ($m_N$) masses in lattice units for 
quenched GDWF with 2-flavor Wilson fermions, $V=16^3 \times 32$, $L_s=16$ 
and  $\m0=1.9$. The $\mf=0.0$ data are from linear extrapolation.
Plotted in Figure 1.
} 
\label{tab_rho_nuc}
\end{table}

\begin{table}
\centering
\begin{tabular}[h]{||c|c||c|c||c|c||c||} \hline \hline
$\beta$	DWF &  $\beta$	GDWF & $m_\rho$ DWF & $m_\rho$ GDWF & $m_N$ DWF & $m_N$ GDWF & $\ainv \gev$    \\ \hline \hline
5.7         &  4.4           & 0.756(22)    & 0.720(12)     & 1.03(6)   & 0.969(23)  & 1.070(18)       \\ \hline
5.85        &  4.6           & 0.549(14)    & 0.541(9)      & 0.74(2)   & 0.712(25)  & 1.423(25)       \\ \hline
6.0         &  4.8           & 0.404(8)     & 0.399(16)     & 0.566(21) & 0.557(30)  & 1.930(77)       \\ \hline \hline
\end{tabular}
\caption{
Matching the scale of quenched GDWF with the scale of
quenched DWF. The $m_\rho$ and $m_N$ data are from $\mf=0$ 
extrapolations and are in lattice units.
The $\ainv$ is derived from the GDWF $m_\rho$ mass.
The DWF data are from reference \cite{columbia_quenched_dwf}.
The parameters are as in table \ref{tab_rho_nuc}.
Plotted in Figures 2 and 3.
} 
\label{tab_match}
\end{table}

\begin{table}
\centering
\begin{tabular}[h]{||c|c|c||c||c||} \hline \hline
$\ainv \gev$  &  $m_\pi \gev$  &  $m_\rho \gev$  &  $m_N \gev$  &   $<\Psibar \gamma_5 \Psi>$  \\ \hline \hline
1.070(18)     &  1.373(25)     &  1.530(26)      &  2.97(24)    &   -0.59(13.77) $10^{-6}$     \\ \hline
1.423(25)     &  1.791(31)     &  1.906(33)      &  3.80(15)    &   -1.65(10.65) $10^{-6}$     \\ \hline
1.930(77)     &  2.318(93)     &  2.406(97)      &  4.71(23)    &   -3.16(11.94) $10^{-6}$     \\ \hline \hline
\end{tabular}
\caption{
The $m_\pi$, $m_\rho$ and $m_N$ in $\gev$ measured using 2-flavor
dynamical Wilson fermions (both for the propagator and the sea quarks)
with mass $\mu=-1.9$ for three values of the lattice spacing.  The
$\ainv$ is derived from the GDWF $m_\rho$ mass and is the same as in
table \ref{tab_match}. All hadron masses are above the cutoff. Also in
the table the expectation value of the Wilson fermion $\Psibar
\gamma_5 \Psi$ condensate is shown in lattice units. It is consistent
with zero well within the error. 
Plotted in Figures 6 and 7.
} 
\label{tab_wilson}
\end{table}

\begin{table}
\centering
\begin{tabular}[h]{||c||c|c||} \hline \hline
\multicolumn{3}{||c||}{
$\ainv = 1.070(18) \gev$}  \\ \hline \hline
$L_s$  &  $\mres {\rm DWF}$  &  $\mres {\rm GDWF}$ \\ \hline \hline
8      &  0.034860(885)      &  0.024820(705)      \\ \hline
16     &  0.016765(610)      &  0.006180(285)      \\ \hline
24     &  0.010395(545)      &  0.002215(120)      \\ \hline
32     &  0.007595(585)      &  0.001075(45)       \\ \hline
40     &  0.006565(270)      &  0.000550(30)       \\ \hline \hline
\end{tabular}
\caption{
The residual mass $\mres$ for quenched DWF and GDWF for various $L_s$ values. 
The parameters are $V=16^3 \times 32$, $\beta_{\rm DWF} = 5.7$, 
$\beta_{\rm GDWF} = 4.4$, $\m0=1.9$, and $\mf = 0.02$.
Plotted in Figure 8.
} 
\label{tab_mres_ls_4.4}
\end{table}

\begin{table}
\centering
\begin{tabular}[h]{||c||c|c||} \hline \hline
\multicolumn{3}{||c||}{
$\ainv = 1.423(25) \gev$}  \\ \hline \hline
$L_s$  &  $\mres {\rm DWF}$  &  $\mres {\rm GDWF}$ \\ \hline \hline
8      &  0.013650(325)      &  0.012700(450)      \\ \hline
12     &  0.006950(385)      &  0.004785(115)      \\ \hline
16     &  0.004585(155)      &  0.002095(65)       \\ \hline
20     &  0.003255(110)      &  0.001035(75)       \\ \hline
24     &  0.002660(100)      &  0.000595(34)       \\ \hline
32     &  0.001950(75)       &  0.000231(14)       \\ \hline
40     &  0.001560(65)       &  0.000102(7)        \\ \hline
48     &  0.001380(60)       &  0.000051(4)        \\ \hline \hline
\end{tabular}
\caption{
The residual mass $\mres$ for quenched DWF and GDWF for various $L_s$ values. 
The parameters are $V=16^3 \times 32$, $\beta_{\rm DWF} = 5.85$, 
$\beta_{\rm GDWF} = 4.6$, $\m0=1.9$, and $\mf = 0.02$.
Plotted in Figure 9.
} 
\label{tab_mres_ls_4.6}
\end{table}

\begin{table}
\centering
\begin{tabular}[h]{||c||c|c||} \hline \hline
\multicolumn{3}{||c||}{
$\ainv =  1.930(77) \gev$}  \\ \hline \hline
$L_s$  &  $\mres {\rm DWF}$  &  $\mres {\rm GDWF}$ \\ \hline \hline
8      &  0.005810(345)      &  0.006640(405)      \\ \hline
16     &  0.001085(70)       &  0.000650(30)       \\ \hline
24     &  0.000495(35)       &  0.0000960(85)      \\ \hline
32     &  0.000310(25)       &  0.00002055(195)    \\ \hline
40     &  0.000225(25)       &  0.00000525(45)     \\ \hline \hline
\end{tabular}
\caption{
The residual mass $\mres$ for quenched DWF and GDWF for various $L_s$ values. 
The parameters are $V=16^3 \times 32$, $\beta_{\rm DWF} = 6.0$, 
$\beta_{\rm GDWF} = 4.8$, $\m0=1.9$, and $\mf = 0.02$.
Plotted in Figure 10.
} 
\label{tab_mres_ls_4.8}
\end{table}

\begin{table}
\centering
\begin{tabular}[h]{||c||c|c|c||} \hline \hline
$\mf$	&  $\mres$, $\beta=4.4$  &  $\mres$, $\beta=4.6$  &  $\mres$, $\beta=4.8$  \\ \hline \hline
0.02    &  0.006180(285)         &  0.002095(65)          &  0.000650(30)          \\ \hline
0.04    &  0.005665(145)         &  0.001945(105)         &  0.000655(25)          \\ \hline
0.06    &  0.005720(160)         &  0.001900(80)          &  0.000650(15)          \\ \hline
0.08    &  0.005665(105)         &  0.001995(80)          &  0.000655(15)          \\ \hline \hline
\end{tabular}
\caption{
The quenched GDWF residual mass $\mres$ for various values of $m_f$ and for three values of $\beta$. 
The parameters are $V=16^3 \times 32$, $L_s=16$ and $\m0=1.9$.
Plotted in Figure 11.
} 
\label{tab_mres_mf}
\end{table}

\begin{table}
\centering
\begin{tabular}[h]{||c||c|c|c||} \hline \hline
$\mf$	&  $m_\pi^2$, $\beta=4.4$  &  $m_\pi^2$, $\beta=4.6$  &  $m_\pi^2$, $\beta=4.8$  \\ \hline \hline
0.00    &  0.021(13)               & -0.005(12)               & -0.0001(94)              \\ \hline
0.02    &  0.137(12)               &  0.086(8)                &  0.063(6)                \\ \hline
0.04    &  0.227(9)                &  0.166(15)               &  0.117(9)                \\ \hline
0.06    &  0.343(12)               &  0.259(14)               &  0.182(13)               \\ \hline
0.08    &  0.451(8)                &  0.363(22)               &  0.255(18)               \\ \hline \hline
\end{tabular}
\caption{
The quenched GDWF pion mass squared $m_\pi^2$ for various values of $m_f$ and for three values of $\beta$. 
The $\mf=0$ data are from linear extrapolation. The parameters are $V=16^3 \times 32$, $L_s=16$ and $\m0=1.9$.
Plotted in Figures 12, 13 and 14.
} 
\label{tab_mpi2_mf}
\end{table}

\vfill
\eject

%%%%%%%%%%%%%%%%%%%%%%%%%%%%%%%%%%%%%%% FIGURES %%%%%%%%%%%%%%%%%%%%%%%%%%%
\eject

\clearpage

\figure{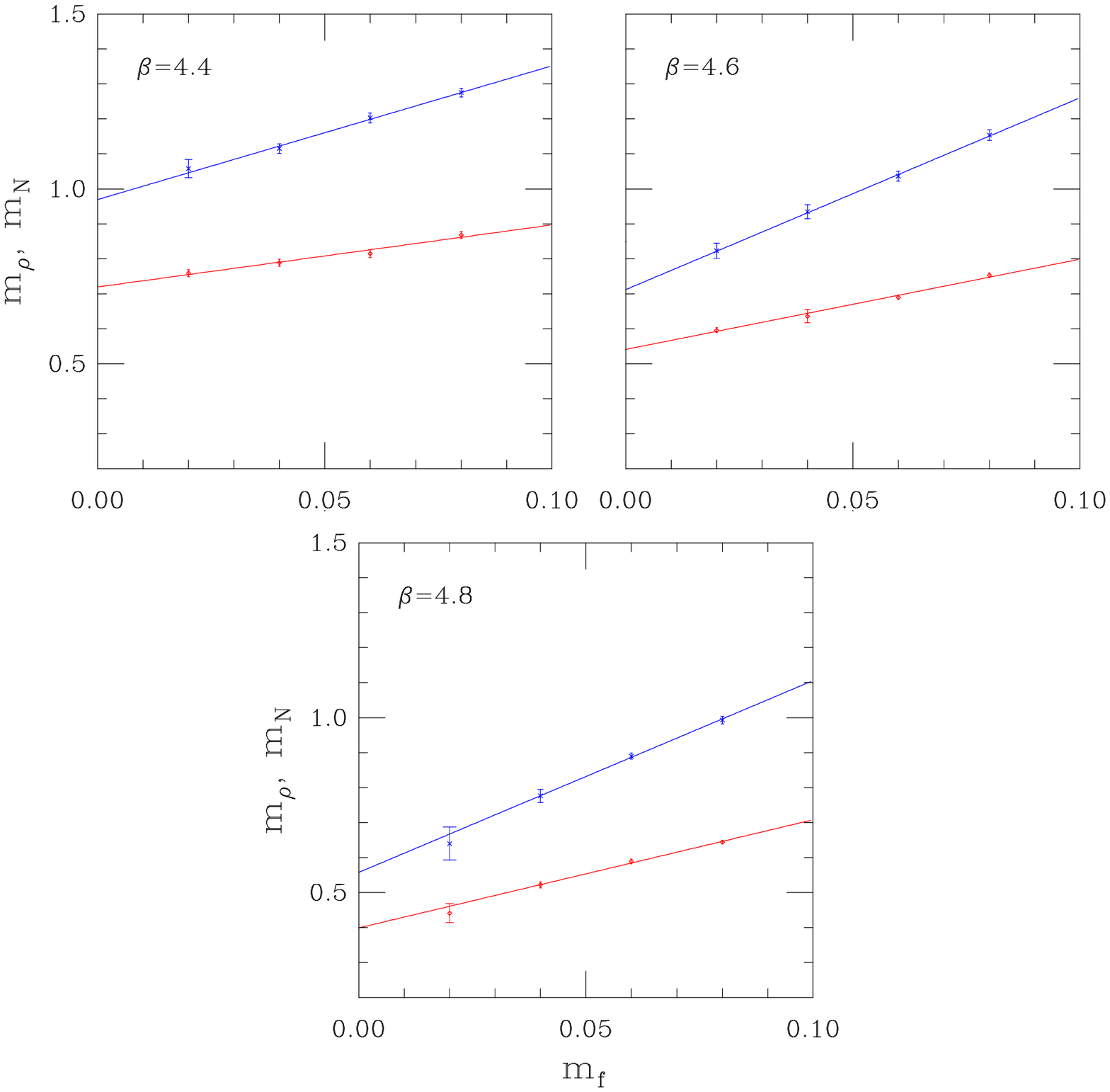}{\one}{\figsizeB}

\figure{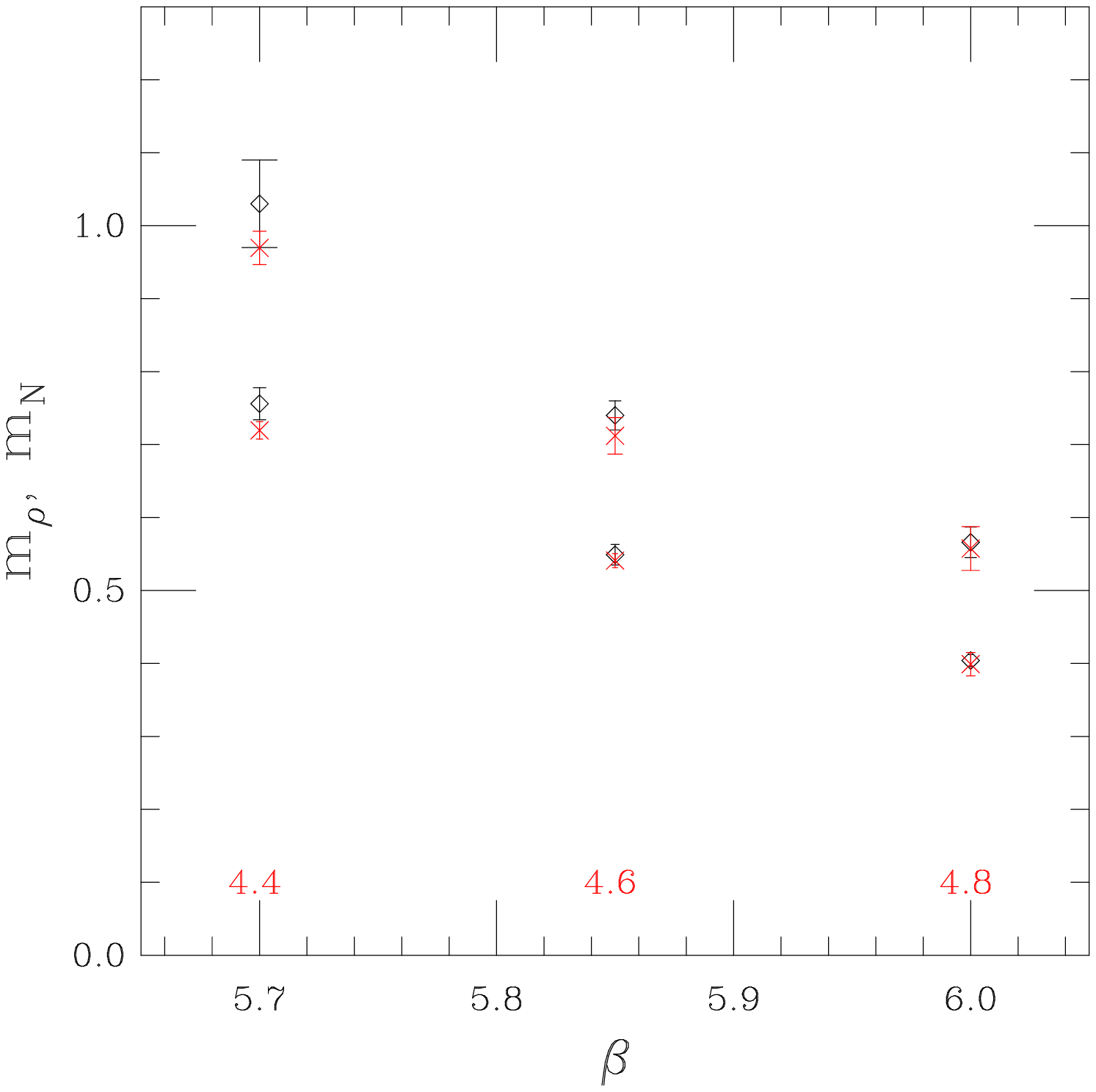}{\two}{\figsizeB}

\figure{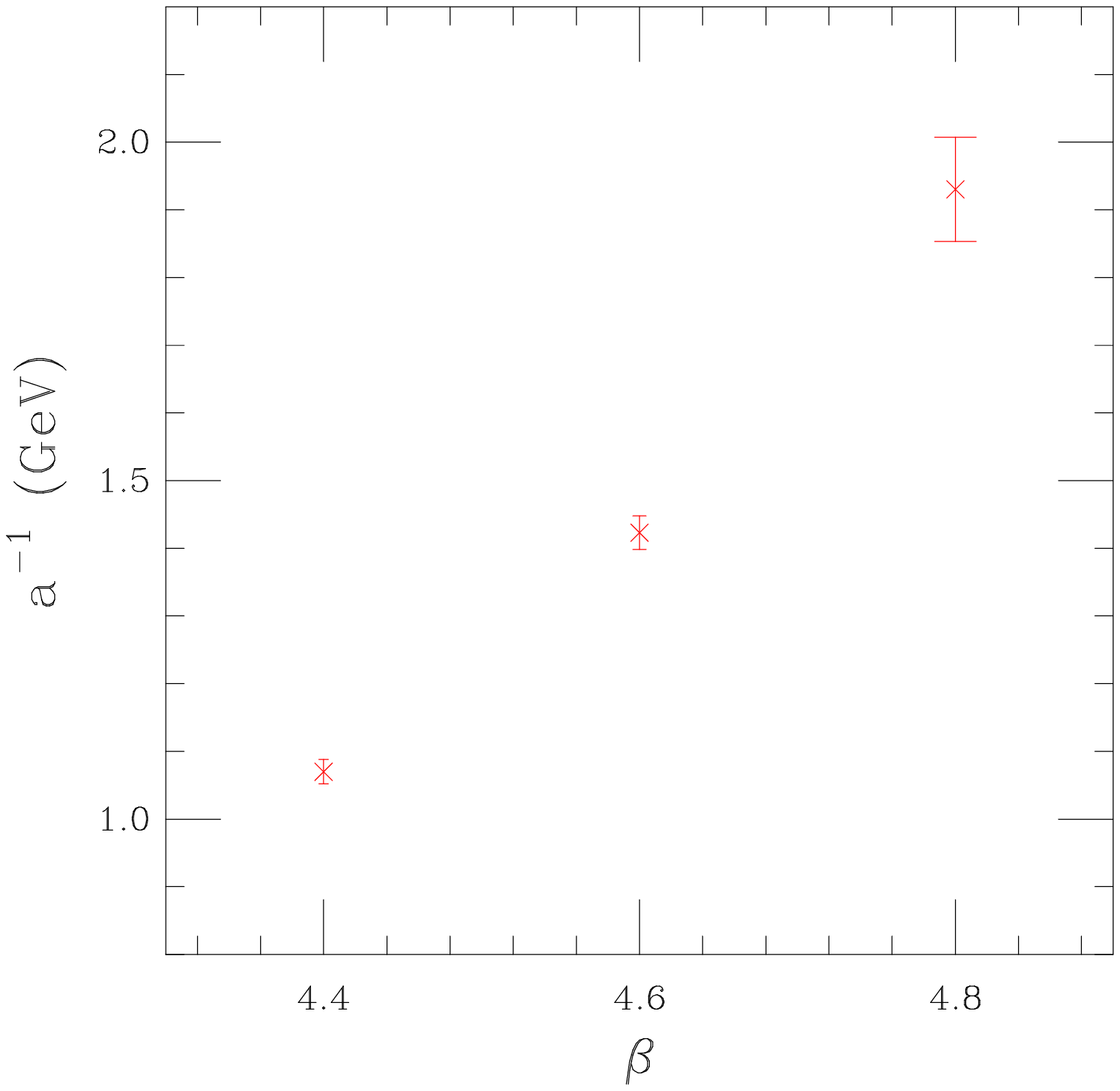}{\three}{\figsizeB}

\figure{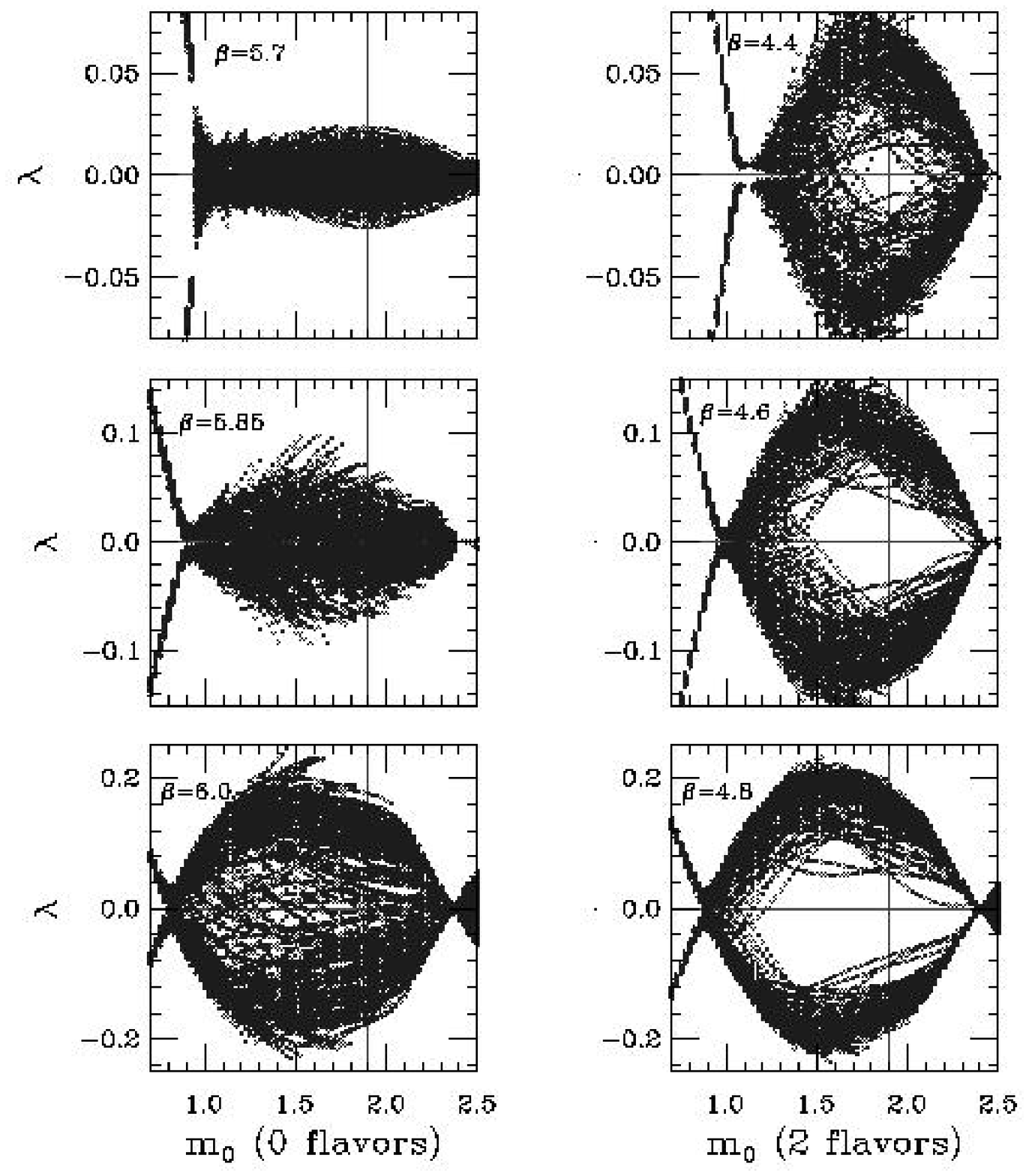}{\four}{\figsizeC}

\figure{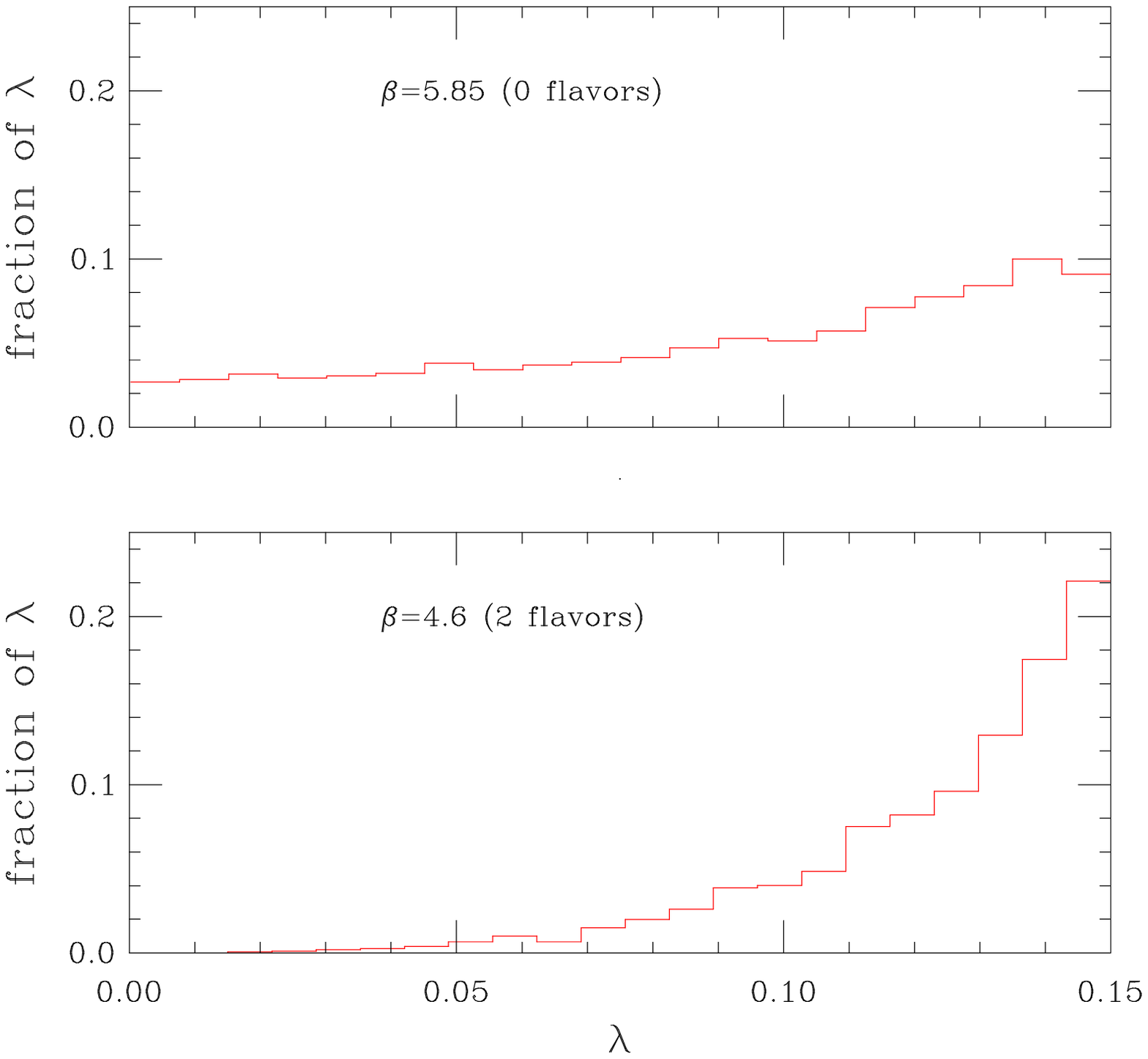}{\five}{\figsizeB}

\figure{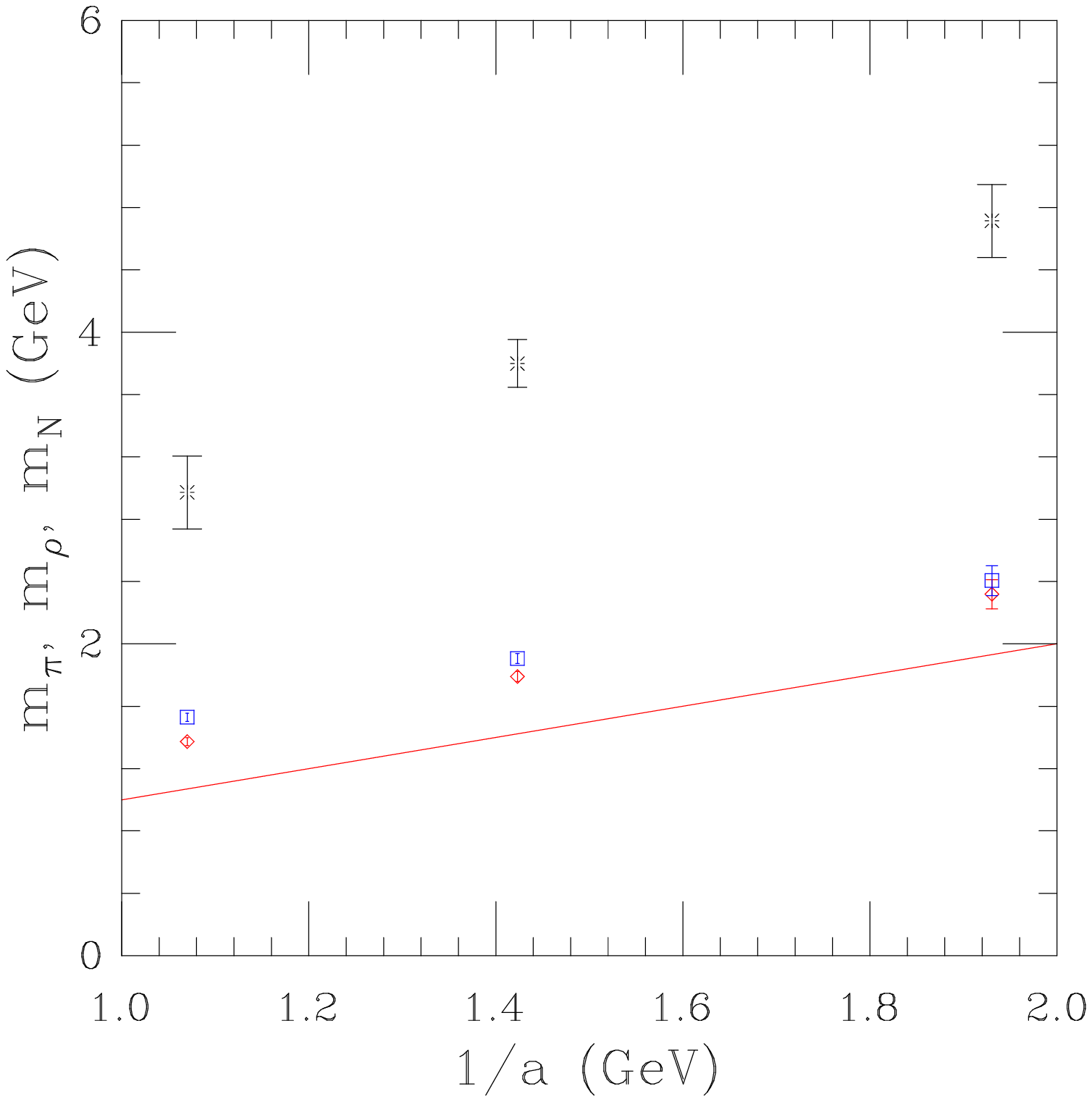}{\six}{\figsizeB}

\figure{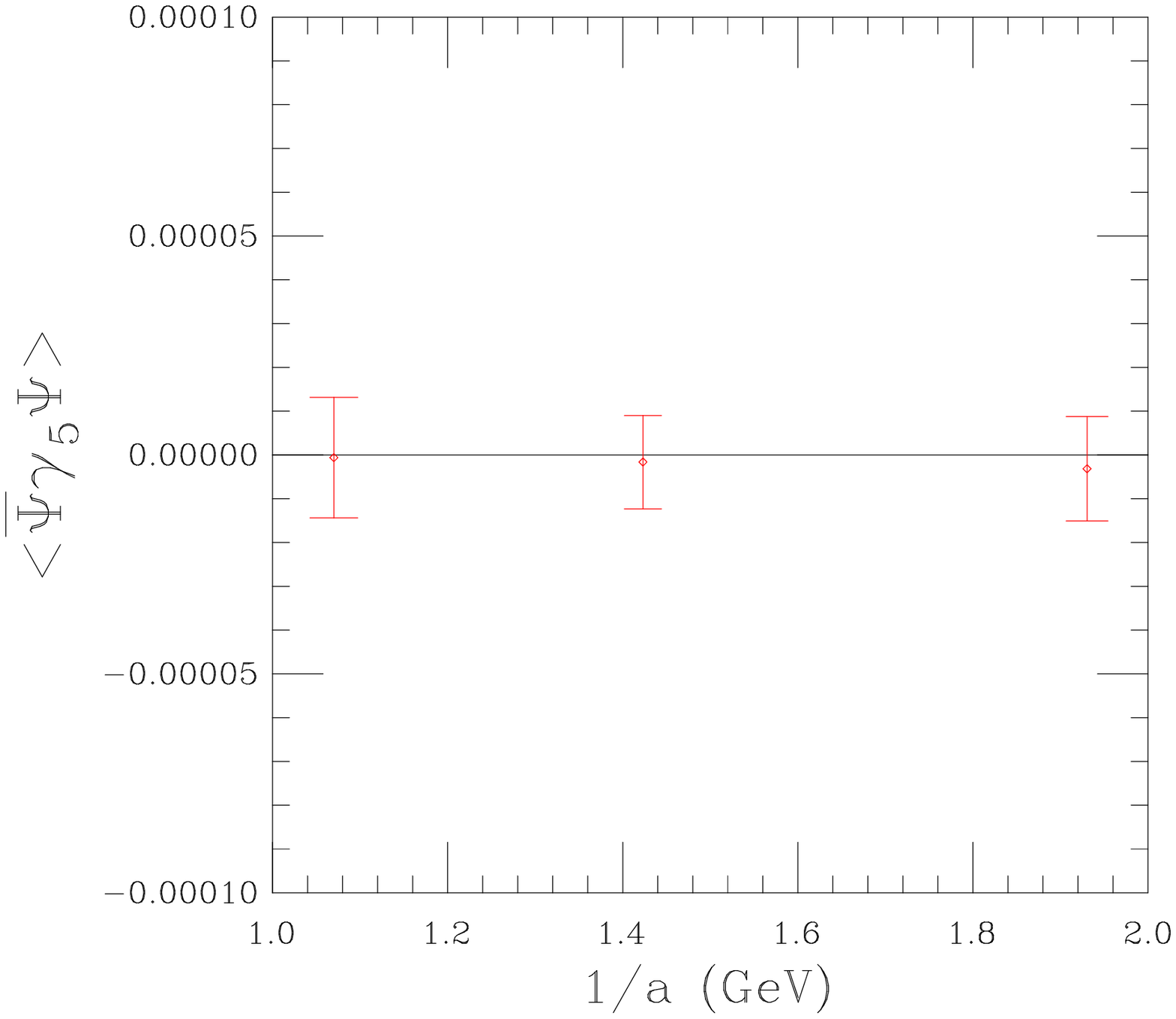}{\seven}{\figsizeB}

\figure{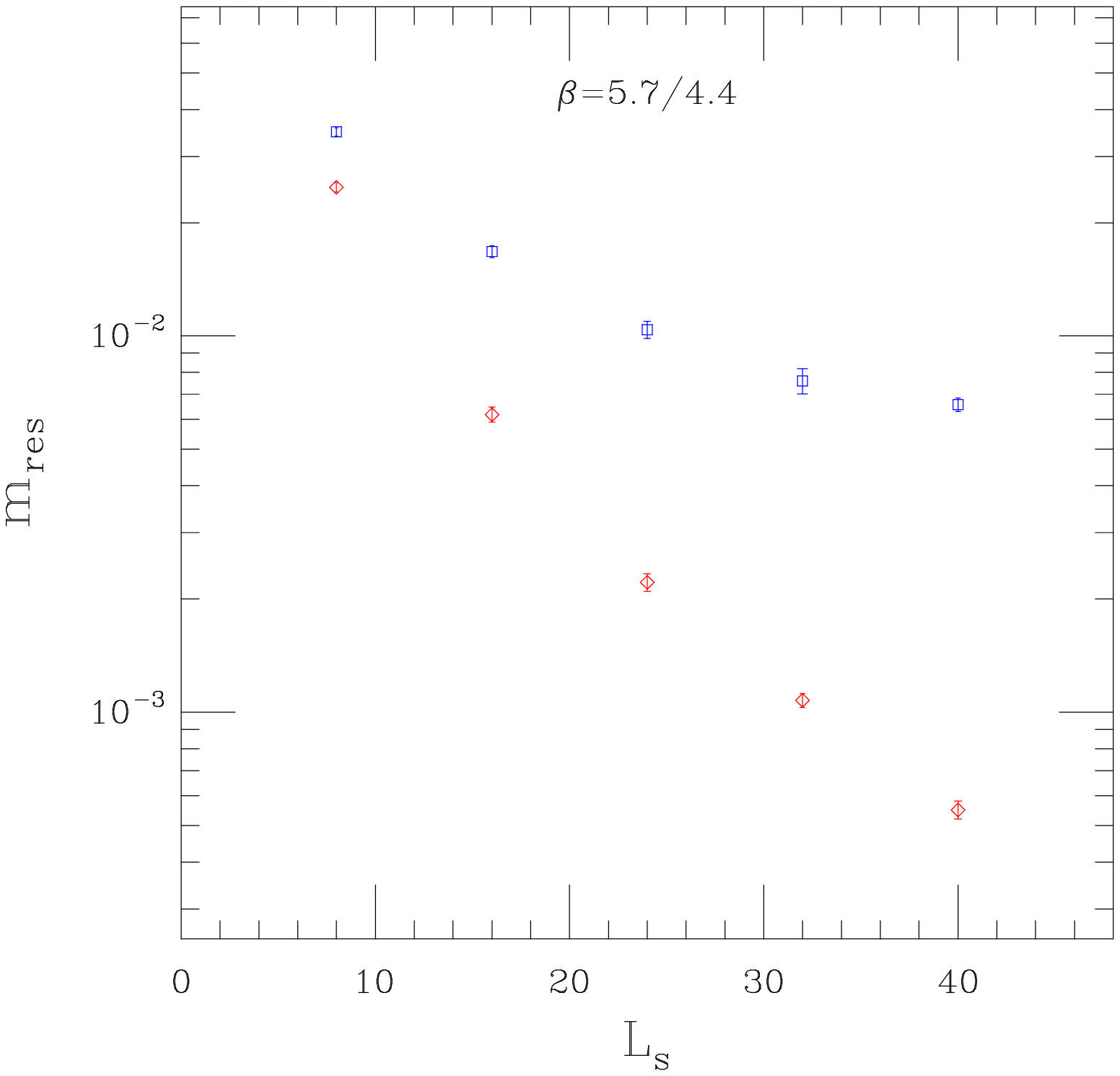}{\eight}{\figsizeB}

\figure{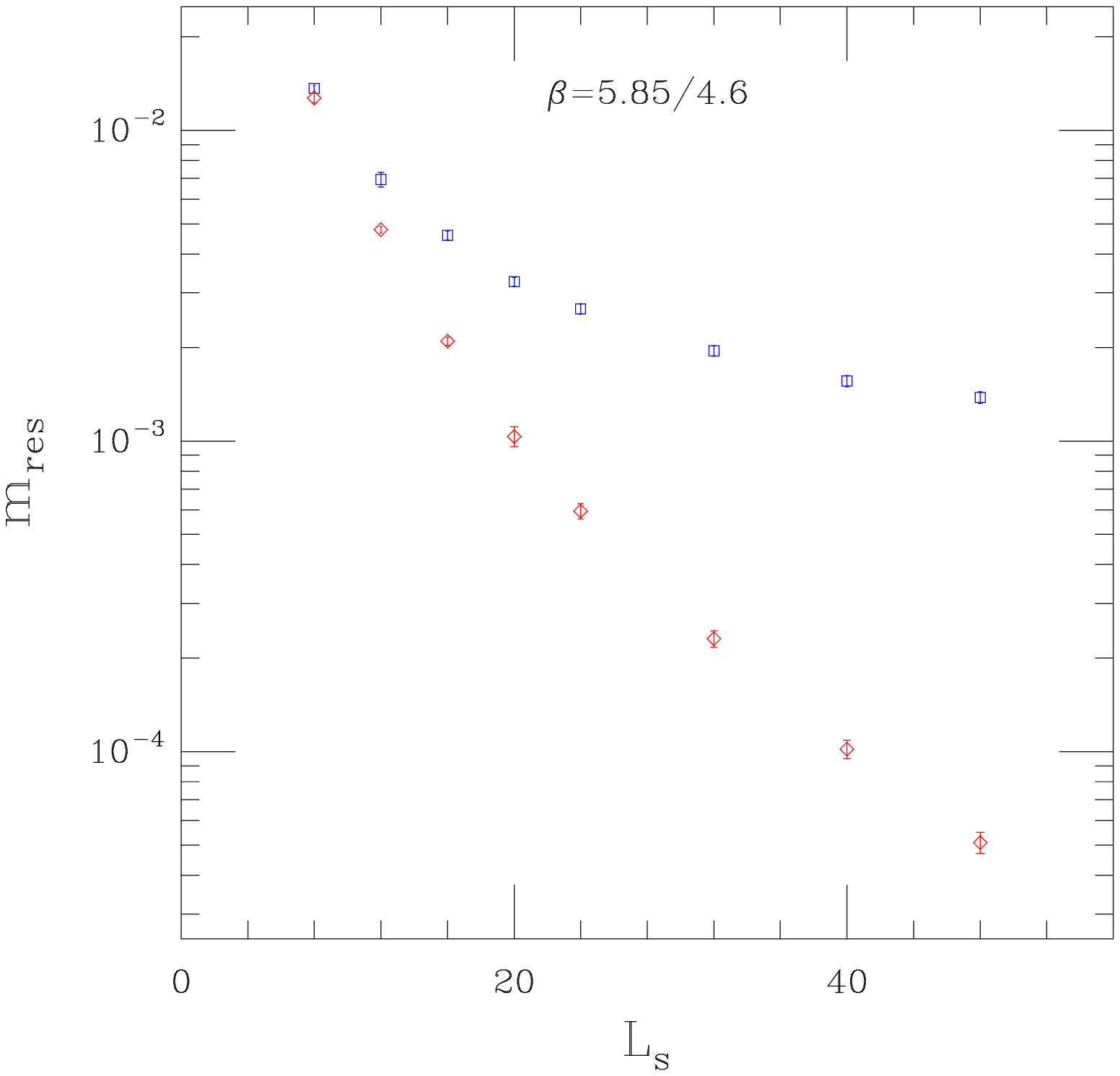}{\nine}{\figsizeB}

\figure{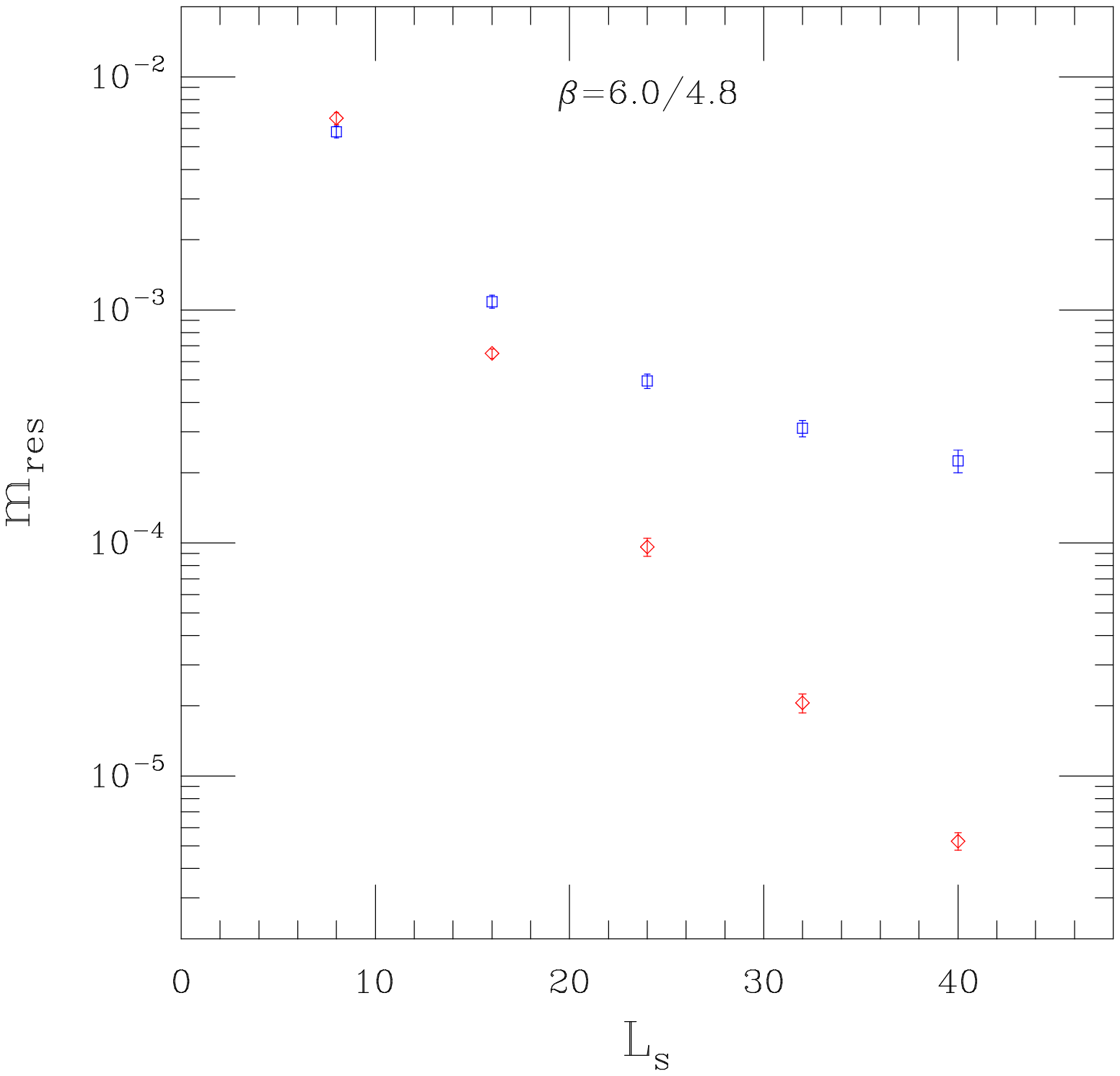}{\ten}{\figsizeB}

\figure{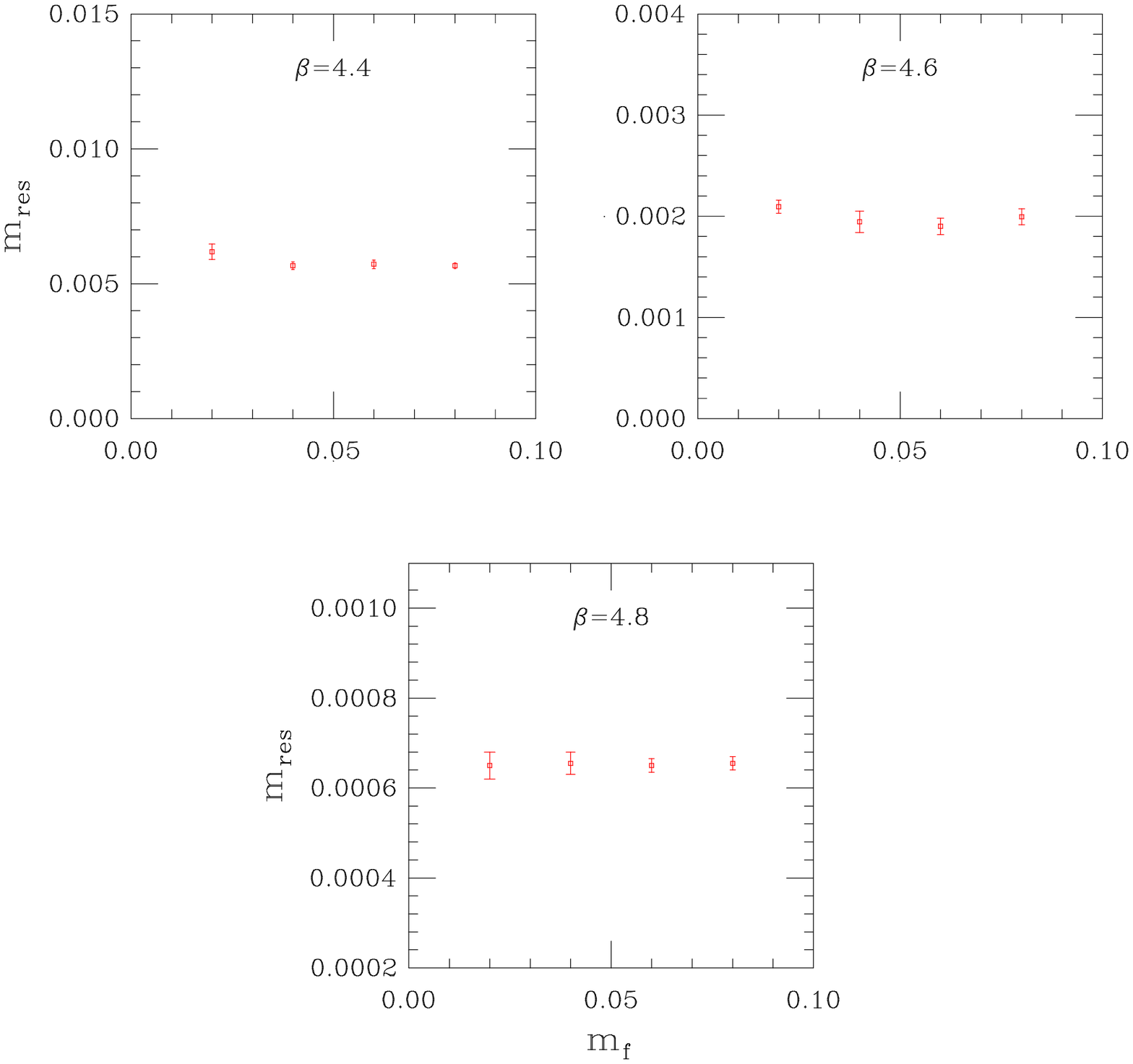}{\eleven}{\figsizeB}

\figure{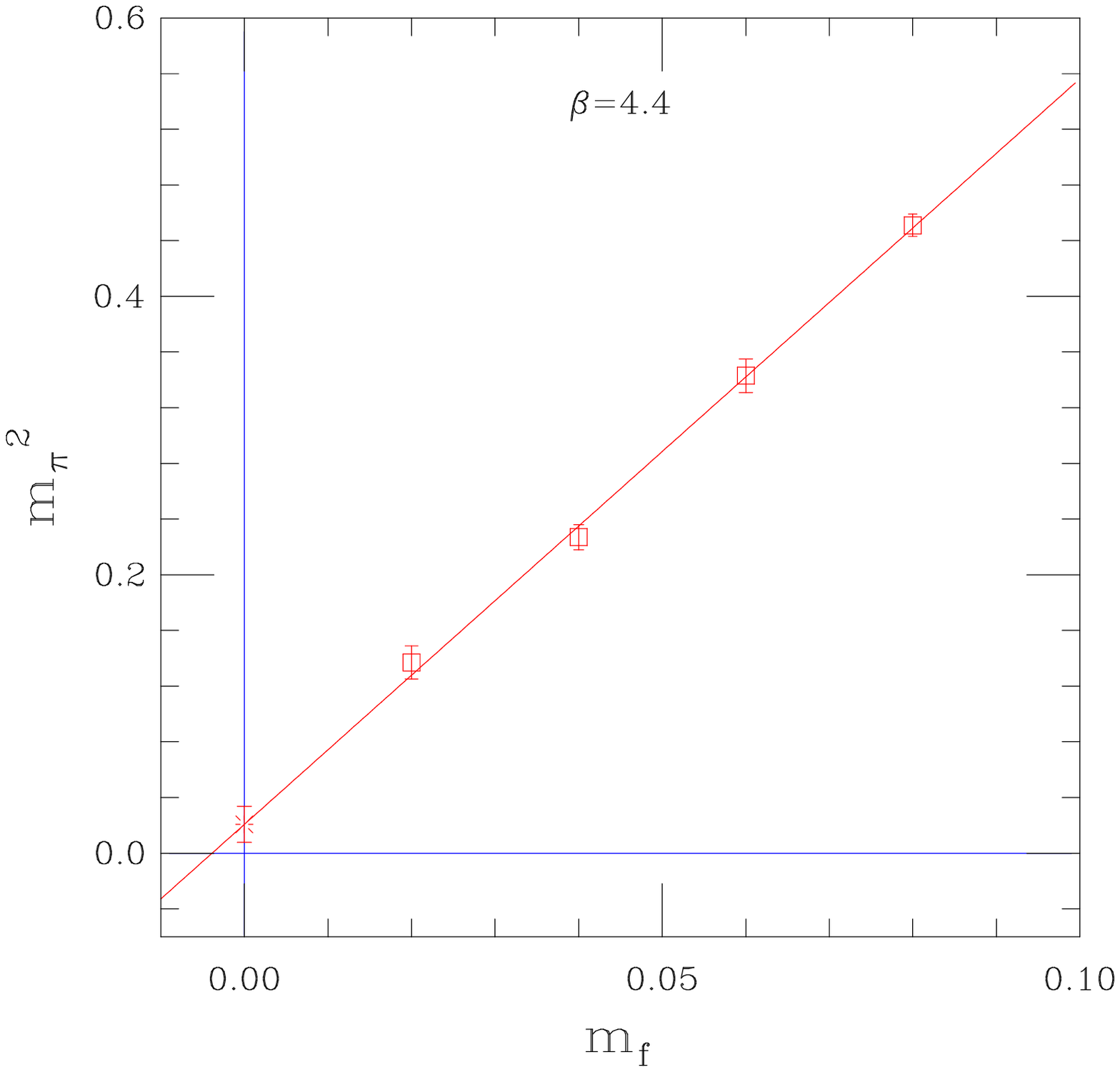}{\twoelve}{\figsizeB}

\figure{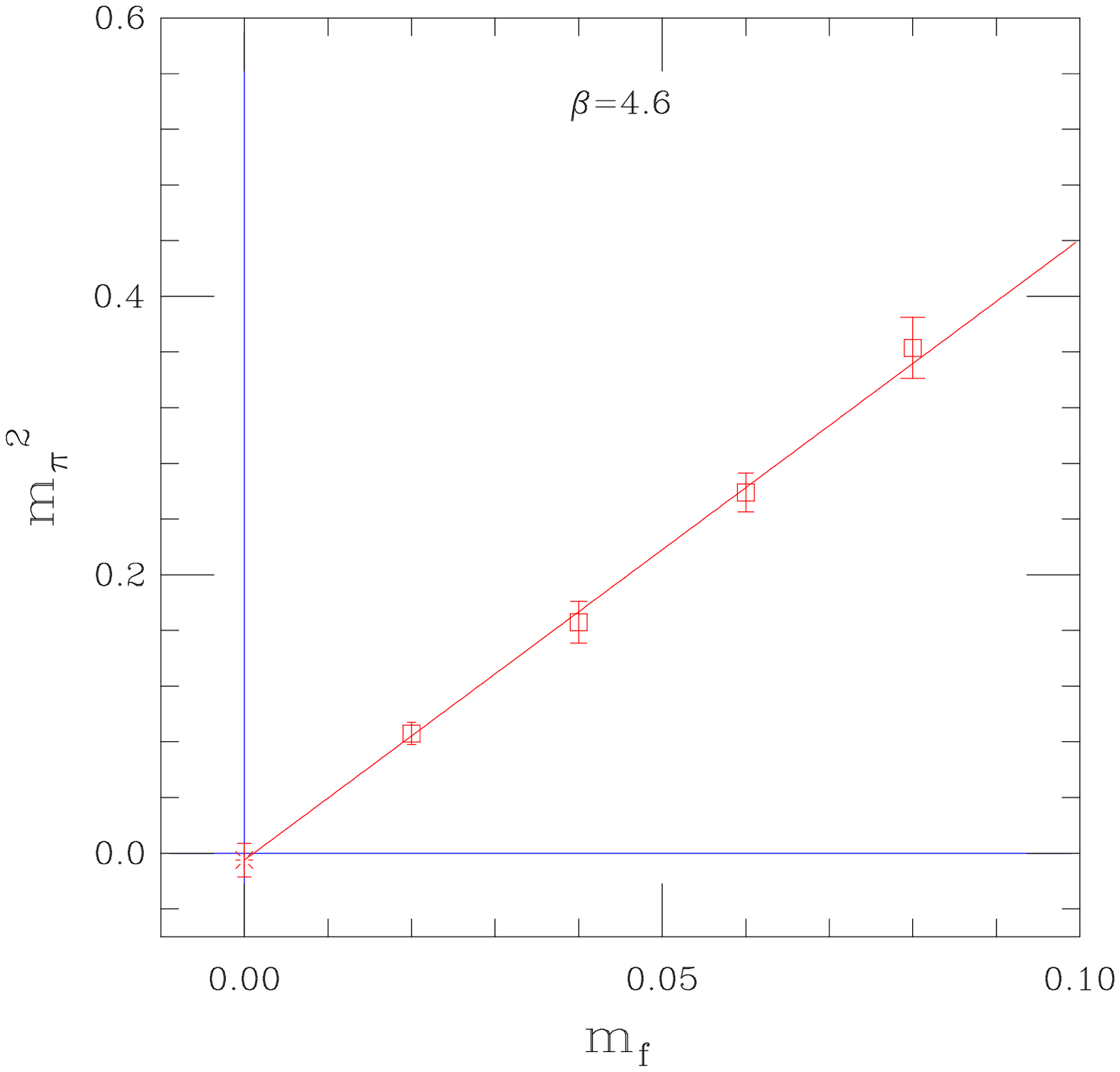}{\thirteen}{\figsizeB}

\figure{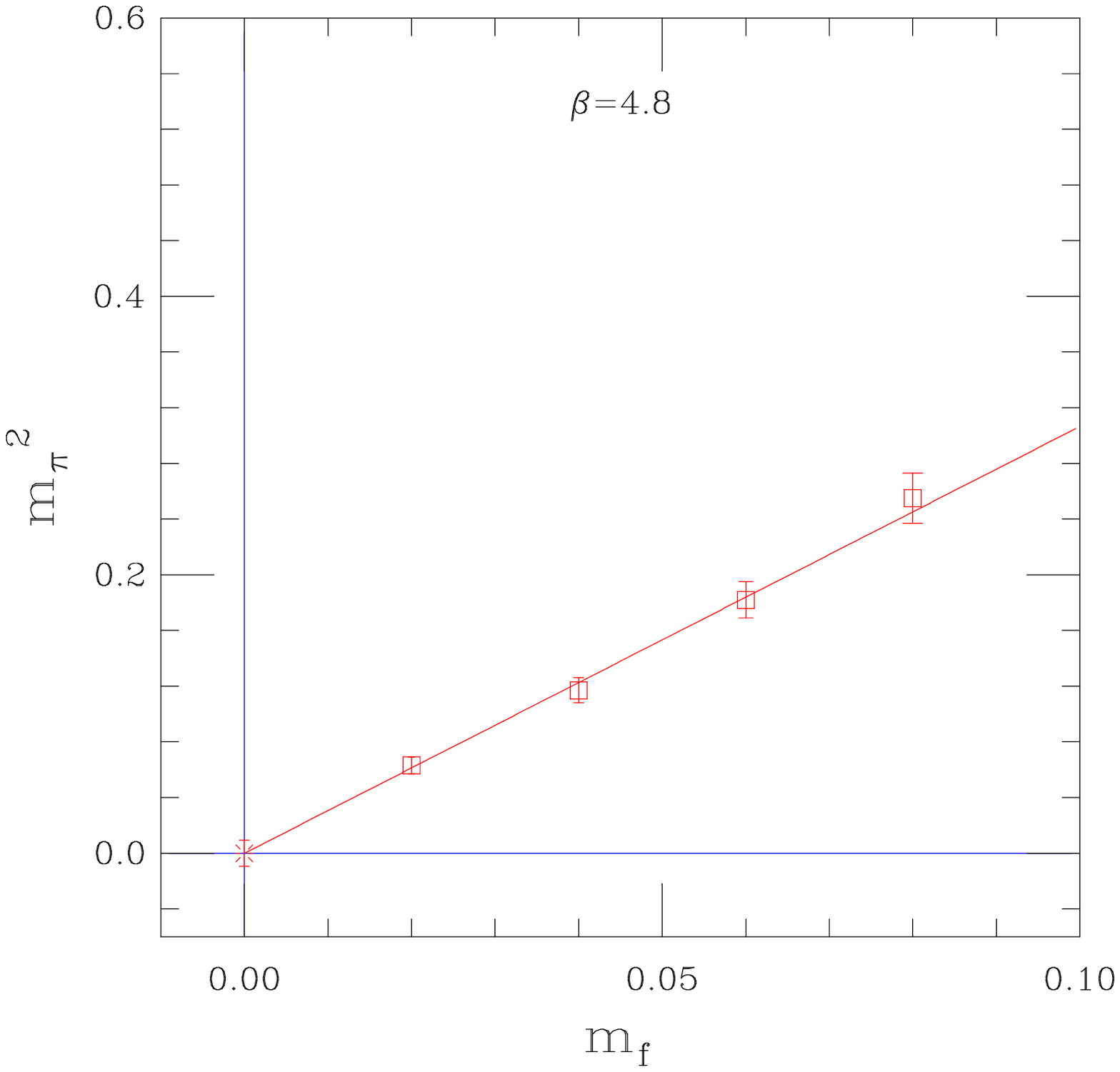}{\fourteen}{\figsizeB}

\end{document}